\documentclass[%
prfluids,%
 amsmath,amssymb,
preprint,%
superscriptaddress]{revtex4-1}

\usepackage{color}
\usepackage{natbib}
\usepackage{graphicx}
\usepackage{dcolumn}
\usepackage{bm}
\usepackage{float}
\preprint{AIP/123-QED}
\newcommand\Rey{\mbox{\textit{Re}}}  
\newcommand\Pran{\mbox{\textit{Pr}}}  
\newcommand\Ha{\mbox{\textit{Ha}}}  
\newcommand\Gras{\mbox{\textit{Gr}}}  
\newcommand\N{\mbox{\textit{N}}}  

\begin{document}

\preprint{APS/123-QED}

\title{Mixed convection in a downward flow in a vertical duct with strong transverse magnetic field}

\author{Xuan Zhang}
\affiliation{Department of Mechanical Engineering, University of Michigan-Dearborn, 48128-1491 MI, USA}
\affiliation{Max Planck Institute for Dynamics and Self-Organization, 37077 G\"ottingen, Germany}
\author{Oleg Zikanov} \email{zikanov@umich.edu}
\affiliation{Department of Mechanical Engineering, University of Michigan-Dearborn, 48128-1491 MI, USA}

\date{\today}

\begin{abstract}
The downward flow in a vertical duct with one heated and three thermally insulated walls is analyzed numerically using the two-dimensional approximation valid in the asymptotic limit of an imposed strong transverse magnetic field.
The work is motivated by the design of liquid metal blankets with poloidal ducts for future nuclear fusion reactors, in which the main component of the magnetic field is perpendicular to the flow direction and very strong heating is applied at the wall facing the reaction chamber. 
The flow is found to be steady-state or oscillating depending on the strengths of the heating and magnetic field. 
A parametric study of the instability leading to the oscillations is performed. It is found among other results that the flow is unstable and develops high-amplitude temperature oscillations at the conditions typical for a fusion reactor blanket.
\end{abstract}

\keywords{Mixed convection, { magnetohydrodynamics}, vertical duct}

\maketitle

\section{Introduction}\label{sec:intro}
Mixed convection in a downward flow in a vertical duct with one wall heated and imposed strong transverse magnetic field is considered (see Fig.~\ref{fig1}). 
The work is motivated by the design of liquid metal blankets for tokamak nuclear fusion reactors. In many design concepts (see, e.g. \cite{Abdou:2015,Smolentsev:2010}), a liquid metal (most likely the PbLi alloy) is pumped through ducts oriented poloidally, i.e. nearly vertically in a large part of the blanket.
 Flow is directed downward in the half of the ducts and upward in the other half. The main component of the strong (4-12 T) imposed magnetic field is perpendicular to the flow direction and to the direction of the strong heat flux from the plasma zone. 

The convection in a downward flow in a vertical round pipe and duct has been investigated experimentally \cite{MelnikovPamir:2014,PamirListratov:2016,Melnikov:2016,Kirillov:2016} and numerically \cite{Liu:2014,Zikanov:2016}. 
It has been found that the thermal convection is a critical factor determining the flow structure. In particular, it leads to the 
large-amplitude low-frequency temperature fluctuations  observed in the experiments \cite{MelnikovPamir:2014,PamirListratov:2016,Melnikov:2016,Kirillov:2016} and reproduced numerically for the pipe flow in \cite{Zikanov:2016}. 

In the flows in pipes and ducts with non-zero mean flow and non-zero wall heating, the mean temperature grows downstream. 
In the case of a vertical duct with a downward flow, the fluid becomes unstably stratified. In an infinite duct, the associated buoyancy force leads to the convection instability in the form of the so-called elevator modes, exact solutions of the governing equations having the form of pairs of ascending/descending vertically uniform jets (see, e.g. \cite{Batchelor:1993,Calzavarini:2006}). In conventional hydrodynamic flows, the jets practically always (with an exception of flows in very narrow tubes \cite{Schmidt:2012}) experience secondary instabilities and break down into turbulence. The situation changes in MHD flows affected by a sufficiently strong magnetic field. As shown for periodic boxes in \cite{Zikanov:2004} and for infinite ducts with transverse field in \cite{Liu:2014}, the jets are stabilized, maintain their vertically uniform shape, and demonstrate exponential growth at very high Grashof numbers.

Numerical simulations of a similar flow in a round pipe of finite length were performed in \cite{Zikanov:2016}. The entire test section of the experiment \cite{Melnikov:2016} was reproduced in high-resolution three-dimensional DNS. Long and strong vertical jets were found to develop, which was attributed to the mechanism similar to that leading to the classical elevator modes and the stabilization by the magnetic field. In a quasi-periodic manner, the jets became unstable and broke down into strong vortices occupying the entire pipe's cross-section and having strong mixing effect. This flow evolution led to strong fluctuations of local temperature. The amplitude and typical frequencies of the fluctuations were in good quantitative agreement with the experimental data of \cite{Melnikov:2016}, which allowed the authors to suggest the growth and breakdown of finite-length quasi-elevator modes as the likely mechanism responsible for the high-amplitude temperature fluctuations in MHD downward flows in vertical pipes and ducts.
 
We should note that the strong effect of thermal convection and development of high-amplitude temperature fluctuations are also observed in other flows with imposed magnetic fields, such as a vertical duct with upward mean flow \cite{Vetcha:2013}, flows with generalized inflectional velocity profiles \cite{Smolentsev:2012,Young:2014}, flows in horizontal ducts and pipes with bottom heating \cite{ZikanovJFM:2013, Zhang:2014, Genin:2011Pamir} or flows in boxes with electrically conducting walls \cite{MistrangeloIEEE:2014,Mistrangelo:2014,Mistrangelo:2013}.
 
For the downward flows in vertical tubes, the previous computational studies were performed either using a simplified model \cite{Liu:2014} or for a round pipe at the Grashof and Hartmann numbers orders of magnitude lower than expected in fusion reactor conditions \cite{Zikanov:2016}. In this paper, we extend the analysis to the rectangular duct geometry (more relevant than the round pipe to the blanket design) and, more importantly, to the wide range of parameters including the values typical for fusion reactors.

\section{Governing Equations and Boundary Conditions}
\label{GE&BC4}
\begin{figure*}
\begin{center}
\includegraphics[width=0.65\textwidth]{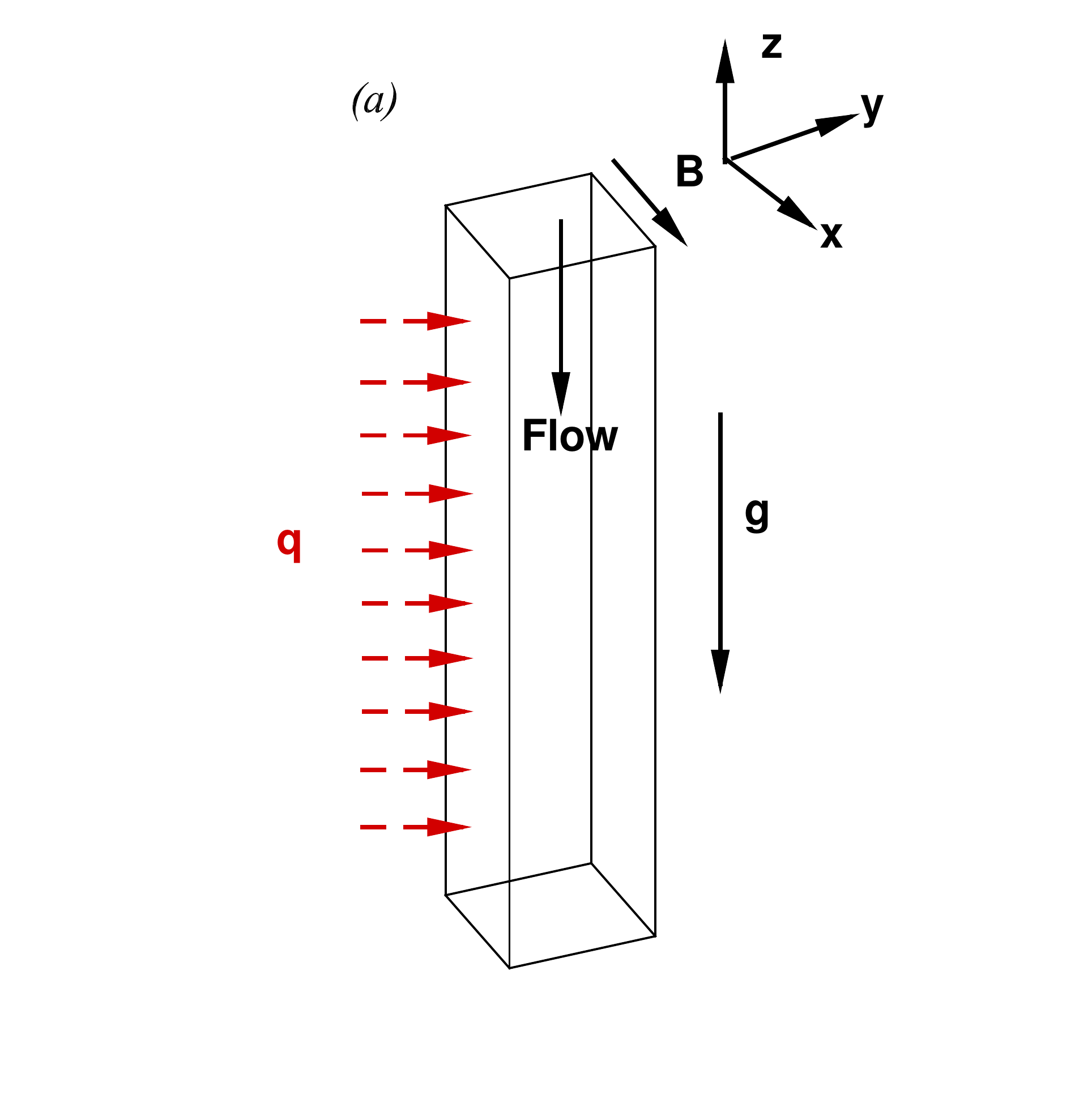}
\includegraphics[width=0.23\textwidth]{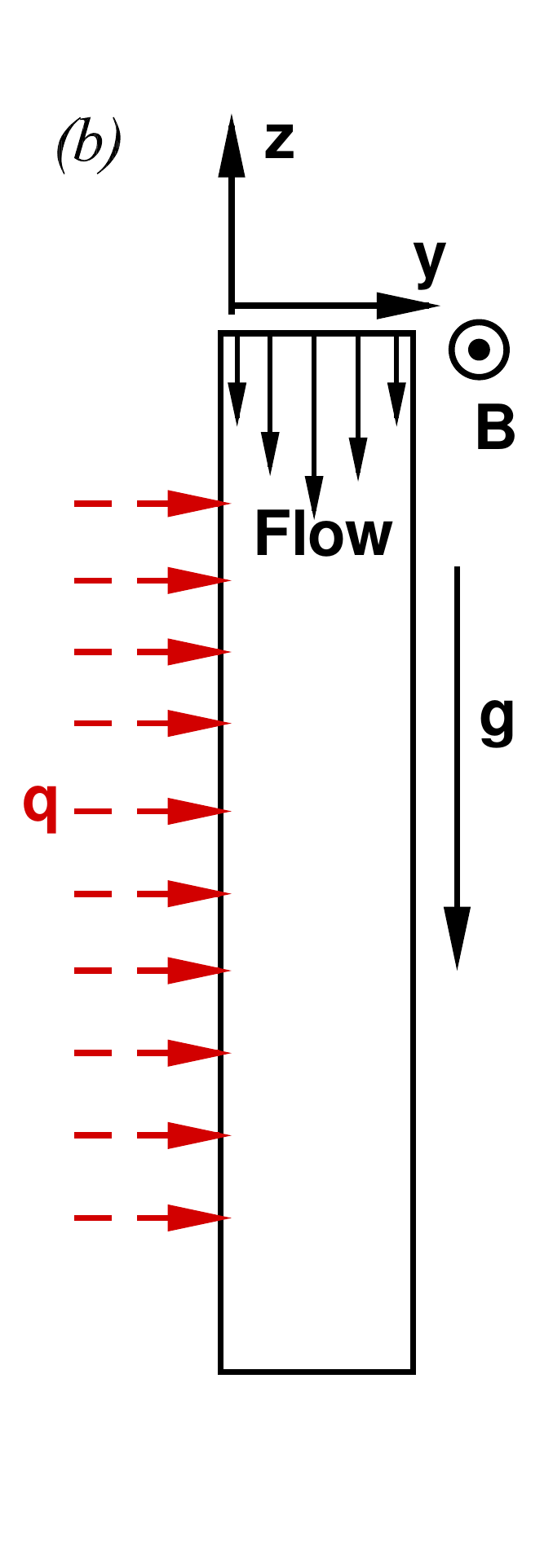}
\caption{Geometry of the flow and the coordinate system for the full three-dimensional flow \emph{(a)}, and the two-dimensional approximation applied in our study \emph{(b)}. $\bm{B}$ is the imposed uniform magnetic field, $\bm{g}$ is the gravity acceleration, $q$ is the uniform surface heating flux. Heating is not imposed in the buffer zones near the inlet and exit.}
\label{fig1}
\end{center}
\end{figure*}

Flows of an incompressible electrically conducting fluid (e.g. a liquid metal) in a long vertical duct of square cross-section are considered (see Fig.~\ref{fig1}). The mean velocity is directed downward and has the constant value $U$. The walls of the duct are electrically perfectly insulated. The uniform  magnetic field of strength $B$ oriented horizontally and perpendicularly to one set of the walls is imposed in the entire flow domain. 

Three walls of the duct are thermally insulated. At the fourth wall, which is parallel to the magnetic field, heat flux of constant rate $q$ is applied. A similar heating scheme was used in the recent pipe flow studies \cite{Melnikov:2016,Zikanov:2016}. The scheme is not fully consistent with the situation in a fusion reactor blanket, where only a part of the heat flux is deposited on the duct wall nearest to the plasma zone. The other, typically larger part is deposited internally via absorption of the neutrons generated in the fusion reaction by the liquid metal. Our reason for  considering wall heating is two-fold. Firstly, the volumetric rate of internal heating in the blanket is strongly concentrated near one wall and decreases exponentially with the distance to it. Secondly, in the existing laboratory experiments, the internal  heating cannot be reproduced. Instead, resistive heating elements are used to apply heat flux at the wall.
Since our work is the first detailed study of the downward  duct flow, we would like to make the model setting closer to the existing experiments, so the results can later be compared with the laboratory data.

Setting the inlet and exit conditions in our configuration is not simple. The ascending and descending jets, which, as we will see, are caused by the buoyancy force, may penetrate inlet and exit. The flow entering a zone of strong magnetic field, especially if this happens, as in a real duct of a fusion reactor blanket, through a manifold, takes a complex three-dimensional form with internal shear layers and, possibly, an M-shaped velocity profile (see, e.g. \cite{Branover:1978} for a discussion and \cite{Li:2013} for an example of numerical analysis). In order to circumvent these aspects and focus the investigation on the effect of magnetoconvection in a long but finite duct, we consider a straight segment of the duct with standard inlet and exit conditions, but introduce buffer zones of the length of 1/6 of the total length of the duct near both inlet and exit. The heating is not applied with these zones. In the presence of strong suppression of velocity non-uniformities by the magnetic field, such buffers are sufficient to minimize the effect of the inlet-exit conditions and render the model closer to the realistic situation. 

The quasi-static approximation of the electromagnetic effects valid for flows with small magnetic Reynolds and Prandtl numbers, which is typical for laboratory and technological flows of liquid metals, is used. The approximation implies that the perturbations of the magnetic field induced by the flow are very weak in comparison with the imposed field, and, therefore, neglected, so the magnetohydrodynamic interactions are reduced to the one-way effect of the field on the flow.

We also assume that the imposed magnetic field is very strong, so the Hartmann and Stuart numbers satisfy
\begin{equation}\label{asl}
\Ha\equiv  Bd\left(\frac{\sigma}{\rho\nu}\right)^{1/2}\gg 1, \; \N\equiv \frac{\sigma B^2 d}{\rho U} \gg 1.
\end{equation}
In these definitions, $d$ is the duct's half-width measured in the direction of the magnetic field, $U$ is the mean velocity, $B$ is the strength of the magnetic field, and $\sigma$, $\nu$, and $\rho$ are the electric conductivity, kinematic viscosity, and density of the fluid. In the asymptotic limit (\ref{asl}), the approximation derived in \cite{Sommeria:Moreau:1982} can be applied to our flow. In the approximation, the magnetic field is assumed to be strong enough to suppress the flow variations along the field lines within the core of the duct. The problem can be approximated as two-dimensional (see Fig.~\ref{fig1}b) expressed in terms of the variables integrated wall-to-wall along the direction of the magnetic field. 
The approximation has been verified and examined in \cite{Potherat:2000,Potherat:2005} and is a common assumption adopted in the previous studies of liquid metal flows in rectangular ducts of a fusion reactor blanket (see, e.g. \cite{Vetcha:2013,Smolentsev:2012,Young:2014}).

The integrated Lorentz force is zero as it should in a flow with perfectly electrically insulating walls. The effect of the magnetic field is reduced to the viscous friction in the thin Hartmann layers, which can be accurately modeled by the linear friction term in the momentum equation. The model was originally developed in \cite{Sommeria:Moreau:1982} for isothermal flows in ducts with electrically insulated walls. It can be easily seen that under the conditions \eqref{asl} and when the imposed heat flux is perpendicular to the magnetic field, the temperature field in a non-isothermal problem can also be accurately approximated by a two-dimensional field.

The governing equations express the conservation of mass, momentum, and energy.  The physical properties are assumed constant with exception of density in the buoyancy force, for which the Boussinesq approximation are applied.
\begin{eqnarray}
\hskip-15mm& &\nabla \cdot\bm{u} = 0,\\
\hskip-15mm& & \frac{\partial \bm{u}}{\partial t} +
(\bm{u}\cdot \nabla)\bm{u}= -\frac{1}{\rho}\nabla P + \nu \nabla^2 \bm{u} - \frac{B}{d}\sqrt{\frac{\sigma\nu}{\rho}}\bm{u} + g\beta T\bm{e}_z,\label{me2}\\
\hskip-15mm& & \frac{\partial T}{\partial t} + (\bm{u}\cdot\nabla)T = \frac{\kappa}{\rho c_p} \nabla^2 T ,
\end{eqnarray}
where $\beta$ is the thermal expansion coefficient, and $\bm{u}$, $P$ and $T$ are wall-to-wall integrated fields of velocity, pressure and temperature deviation from the reference value, for which we take the inlet temperature.
The equations are non-dimesionalized using $d$, $U$ and $qd/\kappa$ as the length, velocity and temperature scales, where $\kappa$ is the thermal conductivity of the fluid. The typical scales of time, pressure and magnetic field are $d/U$, $\rho U^2$ and $B$, respectively.

The non-dimensionalized equations are:
\begin{eqnarray}
\hskip-15mm& &\nabla \cdot\bm{u} = 0,\label{ma}\\
\hskip-15mm& & \frac{\partial \bm{u}}{\partial t} +
(\bm{u}\cdot \nabla)\bm{u}= -\nabla P + \frac{1}{\Rey}\nabla^2 \bm{u} - \frac{\Ha}{\Rey}\bm{u} + \frac{\Gras}{\Rey^2}T \bm{e}_z,
\label{me}\\
\hskip-15mm& & \frac{\partial T}{\partial t} + (\bm{u}\cdot\nabla)T= \frac{1}{\Rey\Pran} \nabla^2 T .\label{en}
\end{eqnarray}

The non-dimensional parameters are the Hartmann number $\Ha$ defined in (\ref{asl}), 
the Reynolds number
\begin{equation}\label{reynolds}
\Rey\equiv \frac{Ud}{\nu},
\end{equation}
the  Prandtl number 
\begin{equation}
\label{prandtl}
\Pran=\nu/\chi,
\end{equation}
and the Grashof number
\begin{equation}\label{grashof}
\Gras\equiv  \frac{g\beta q d^4}{\kappa\nu^2},
\end{equation}
where $\chi$ is the temperature diffusivity of the liquid.
The combination $\Ha/\Rey$ represents the effect of magnetic damping through friction in the  Hartmann layers.
The strength of the buoyancy force is determined by the combination $\Gras/\Rey^2$.

The walls are assumed electrically perfectly insulated and no-slip. 
One wall is subjected to constant heat flux (except for the buffer portions near the inlet and exit), and the other wall is thermally insulated: 
\begin{eqnarray}
\hskip-15mm \bm{u}=0 & &\textrm{ at } y=\pm 1,\\
\hskip-15mm \frac{\partial T}{\partial y}= -1 & &\textrm{ at } y= -1,\, \frac{L_z}{6}\le z \le \frac{5L_z}{6},\\
\hskip-15mm \frac{\partial T}{\partial y}= 0 & &\textrm{ at } y= -1,\, z < \frac{L_z}{6}, z>\frac{5L_z}{6},\\
\hskip-15mm \frac{\partial T}{\partial y}= 0 & &\textrm{ at } y=1.
\end{eqnarray}
The parabolic velocity profile and isothermal flow are imposed at the inlet. 
Conditions of free flow are applied at the exit: 
\begin{eqnarray}
\hskip-15mm u_y = 0, \, u_z = 1.5(1-y)^2, \;T = 0 & &\textrm{ at } z= 0,\\
\hskip-15mm \frac{\partial u_y}{\partial z} = \frac{\partial u_z}{\partial z} = \frac{\partial T}{\partial z}= 0 & &\textrm{ at }  z= L_z.
\end{eqnarray}
Here, $L_z$ is the non-dimensional duct length.

Small-amplitude random perturbations of velocity and temperature distributed around zero are added to initialize the flow.

\section{Parameters, Numerical Method and Computational Grid}
\label{grid4}
The Prandtl number is fixed at $\Pr=0.0321$, which corresponds to the LiPb alloy at 570 K \cite{Schulz:1991}.  As a model of flows in dual-coolant and self-cooled liquid metal blankets \cite{Abdou:2015}, we study flows at $5000 \le\Rey\le 10^6$,  $10^{6}\le \Gras \le 10^{11}$ and $10^3\le \Ha\le 10^4$.
The non-dimensional duct length is $L_z=30$ or $L_z=60$. 

The problem is solved numerically using the second-order finite-difference scheme developed for magnetohydrodynamic flows in \cite{Krasnov:FD:2011} and later extended to flows with thermal convection in \cite{ZikanovJFM:2013}. The method has demonstrated its efficiency and accuracy in simulations of liquid metal flows at high values of $\Ha$, $\Rey$ and $\Gras$ (see, e.g. \cite{Krasnov:2012,Zhang:2014,Zhang:2015,Zhang_P1:2016,Zhang_P2:2016}). The key feature of the method is the highly conservative approximation. In the non-diffusive limit, the discretized model conserves mass, momentum, internal energy, and electric charge exactly, while the kinetic energy is conserved with the dissipative error of the third order. Further information can be found in the references just mentioned.

The computational grid is structured Cartesian with the points distributed uniformly in the $z$-direction and clustered toward the walls according to the coordinate transformation:
\begin{equation}\label{tanh} 
y=\frac{\tanh\left(A_y \eta\right)}{\tanh\left(A_y \right)}.
\end{equation}
The stretching coefficient $A_y$ represents the degree of near-wall clustering, and the grid in the transformed coordinate $-1\le \eta \le1$ is uniform. 

The grid sizes and the values of $A_y$ used in the simulations are listed in table \ref{tab1}. They have been determined in the grid sensitivity studies, where we have compared 
 the computed values of the integral characteristics such as the average kinetic energy $E$, integrated square of temperature deviation $E_T$,  mean temperature $\overline{T}$, and the $y$-averaged temperature in the middle of the duct $T_{ave}(L_z/2)$: 
\begin{eqnarray}
E & = & \frac{1}{A}\int_A (|u_y|^2+ |u_z|^2) dA, \label{ke}\\
E_T & = & \frac{1}{A}\int_A T^2 dA, \label{et}\\
\overline{T} & = & \frac{1}{A}\int_AT dA,\label{tm}\\
T_{ave}(L_z/2) & = & \frac{1}{2}\int_{-1} ^{1} T(y,z=L_z/2) dy, \label{tave}
\end{eqnarray}
where A is the area of the flow domain. 

We have found that  the resolution within the Shercliff layers (the magnetohydrodynamic boundary layers of thickness $\sim d\Ha^{-1/2}$ at the duct walls parallel to the magnetic field, i.e. the walls at $y=\pm 1$ in our model) is critical for accurate reproduction of the flow structure. Specifically, the model is accurate when at least 8 grid points are kept across each Shercliff layer.  As an example, the grid steps used in the computations at $\Gras=10^{10}$ and $\Ha=10^4$ are ${\Delta y}_{min} = 0.0015$, ${\Delta y}_{max} = 0.02$ and $\Delta z = 0.025$.

\begin{table}
  \begin{center}
  \begin{tabular}{@{}ccccccc@{}}
  $\Gras$ & $\Ha$ & $\Rey$ & $L_z$ & $N_y$ & $A_y$ & $N_z$  \\
  $10^8$ & $10^3 \le \Ha \le 10^4$ & $5000 \le \Rey \le 10^4$ & 30 & 120 & 1.5 &  1200 \\
  $10^8$ & $10^3 \le \Ha \le 10^4$ & $5000 \le \Rey \le 10^4$  & 60 & 120 & 1.5 & 1800 \\
  $10^8$ & $\Ha=10^4$ & $\Rey \ge 2\times10^4$ & 30  & 120 & 2.0 & 1200 \\
  $10^9$ & $10^3 \le \Ha \le 10^4$ & $ 5000 \le \Rey \le 10^5$ & 30  & 160 & 2.0 & 1200\\ 
  $10^9$ & $10^3 \le \Ha \le 10^4$ & $ 5000 \le \Rey \le 10^5$ & 60  & 160 & 2.0 & 1800\\  
  $10^{10}$ & $\Ha=10^4$ & $ 5\times10^4 \le \Rey \le 10^5$ & 30  & 200 & 2.0 & 1200\\ 
  $10^{11}$ & $\Ha=10^4$ & $ 5\times10^5 \le \Rey \le 10^6$ & 30  & 200 & 2.0 & 1200   
  \end{tabular}
  \end{center}
  \caption{Parameters studied and the corresponding numerical resolution. }
  \label{tab1}
\end{table}

\section{Results}
\label{results4}

As we have already discussed in section \ref{sec:intro}, the heating applied to one of the duct's wall results in unstable density stratification and modification of the flow by the buoyancy force. The numerical simulations presented later in this section indicate that in the parameter range considered in our work the modification is always significant. 

\begin{figure*}
\begin{center}
\includegraphics[width=0.7\textwidth,trim=4 4 4 4,clip]{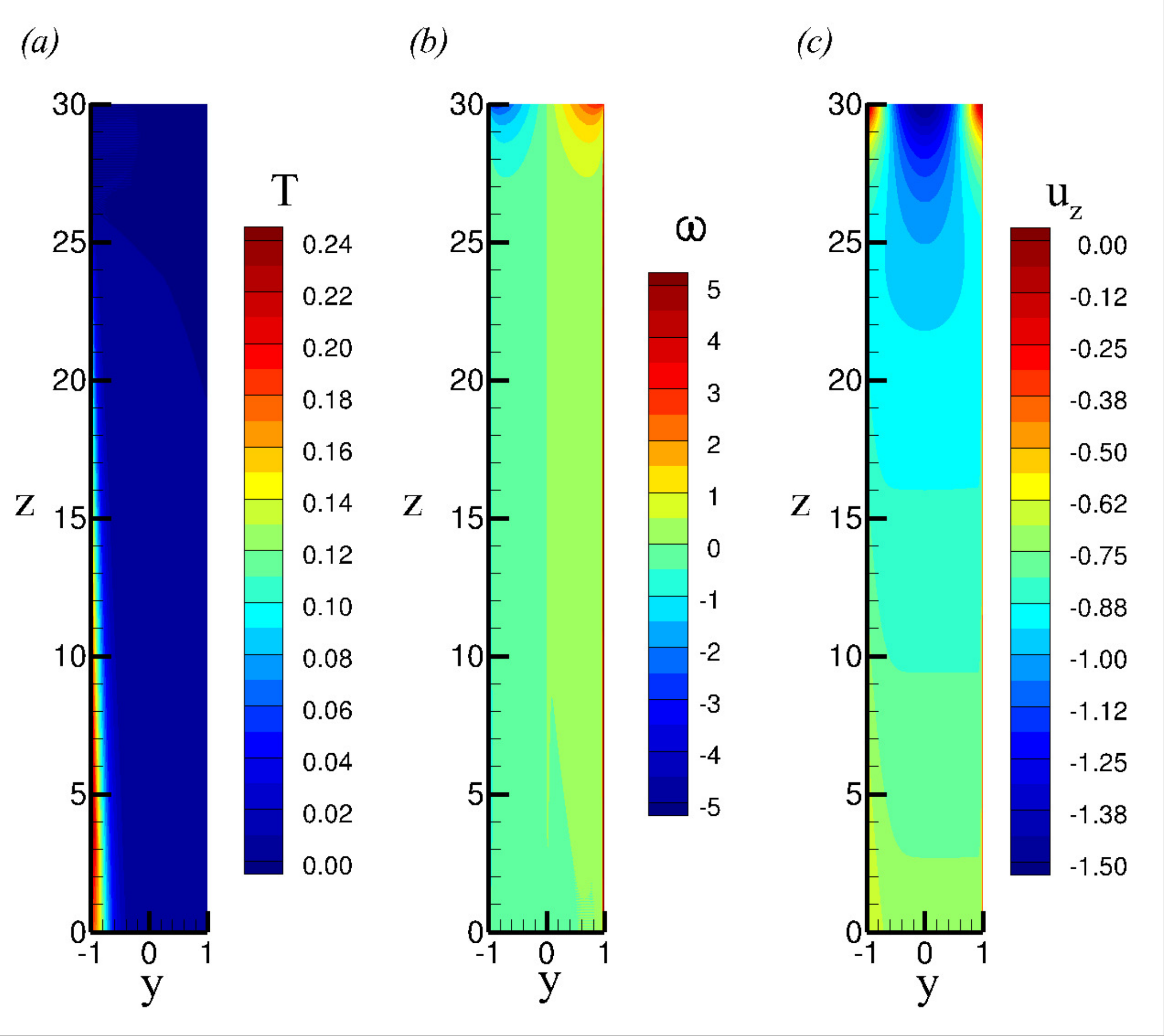}\\
\includegraphics[width=0.7\textwidth,trim=4 4 4 4,clip]{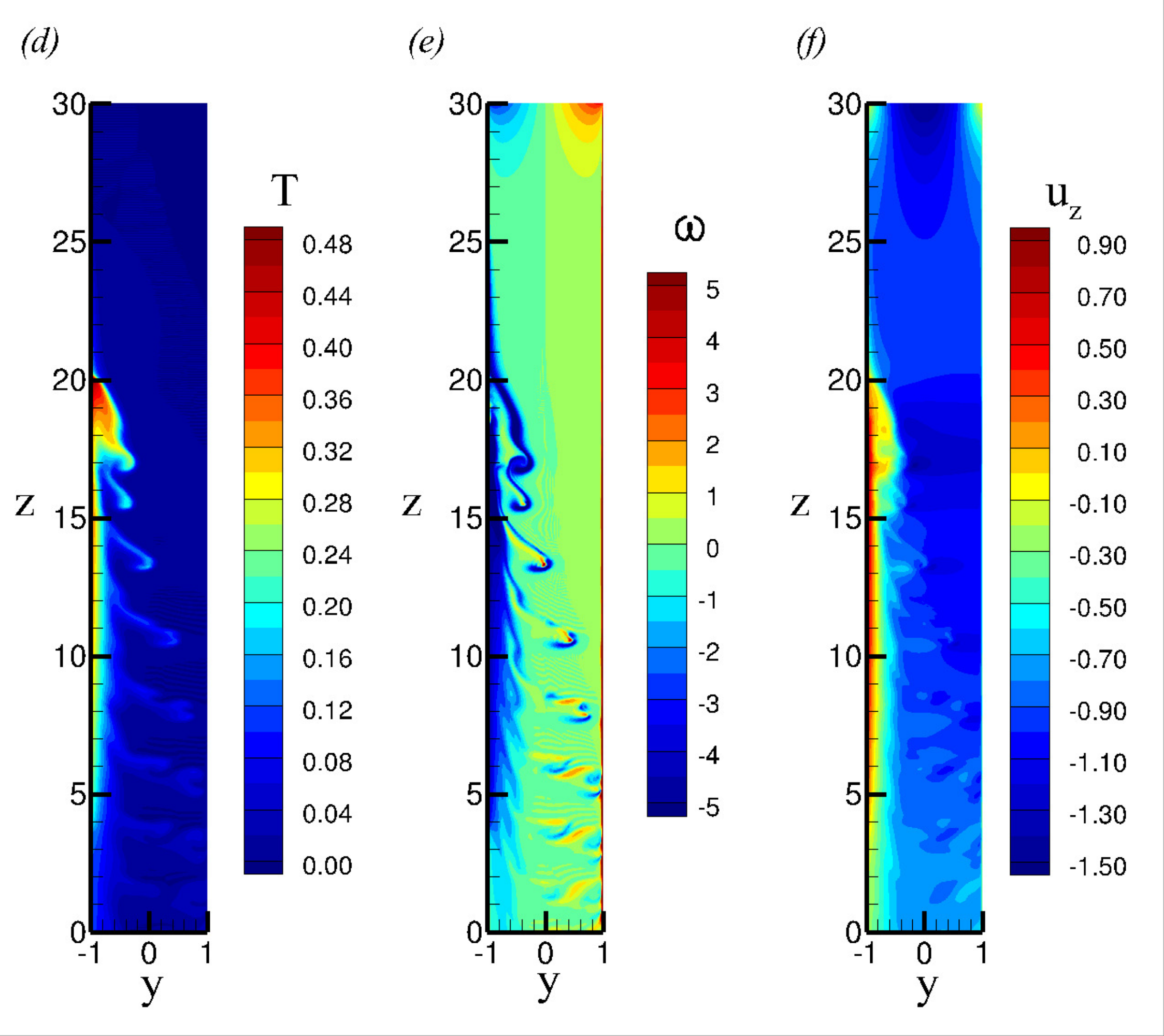}
\caption{Illustration of the stable \emph{(a)-(c)} and unstable \emph{(d)-(f)} states of the flow.  Instantaneous distributions of temperature $T$, vorticity $\omega$ and amplitude of vertical velocity $u_z$ are shown for $\Ha=10^4$, $\Rey=2\times10^4$, $\Gras=10^8$ \emph{(a)-(c)} and $\Gras=10^9$ \emph{(d)-(f)}.  } 
\label{fig2}
\end{center}
\end{figure*}

The two principally different states of the flow found in the simulations are illustrated in Fig.~\ref{fig2}.  One of them is the stable regime shown in Figs.~\ref{fig2}a-c. 
In this flow, a thermal boundary layer develops near the heated wall. The increase of temperature in this layer and the associated buoyancy force are not strong enough to cause the development of a reverse upward jet. The velocity profile, however, becomes asymmetric, with lower downward velocity near the heated wall (see Fig.~\ref{fig2}c). 

In the more interesting case of the strong effect of the buoyancy force, a thin upward jet develops near the heated wall. A much thicker zone of downward velocity is observed across the rest of the duct's width. We have found that this structure is always unstable in the sense that growth of the two-jet pattern inevitably eventually leads to the instability and development of the flow pattern, an example of which is shown in Figs.~\ref{fig2}d-f. We see in Figs.~\ref{fig2}d and~\ref{fig2}e that the instability is initiated at the top of the upward jet. It results in formation of two-dimensional rolls, which grow and penetrate toward the unheated wall as they are transported downward by the mean flow.  It is plausible to assume that the mechanism of this evolution is the Kelvin-Helmholtz instability of the shear layer between the two jets.

\begin{figure*}
\begin{center}
\includegraphics[width=0.9\textwidth]{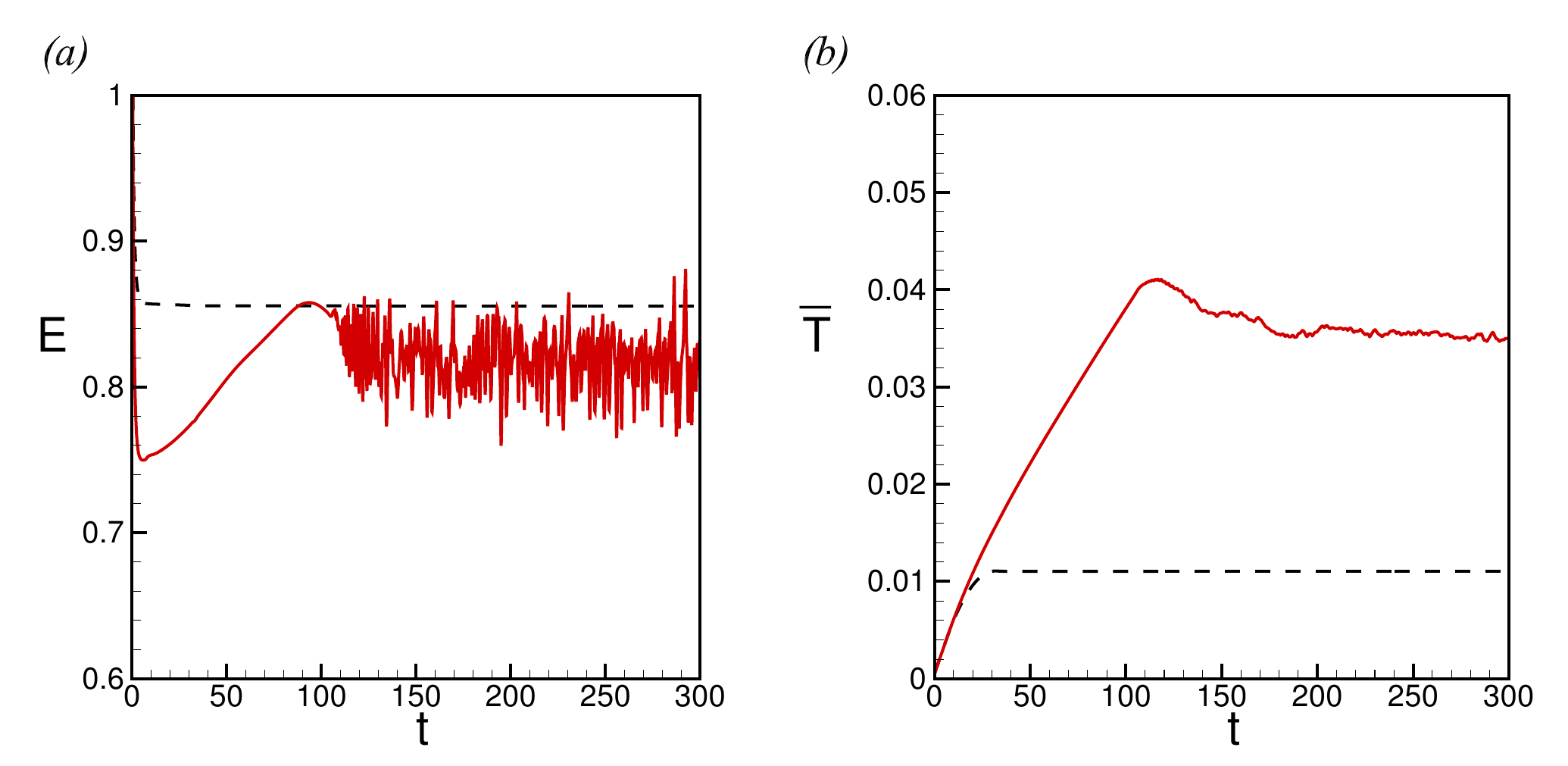}\\
\includegraphics[width=0.9\textwidth]{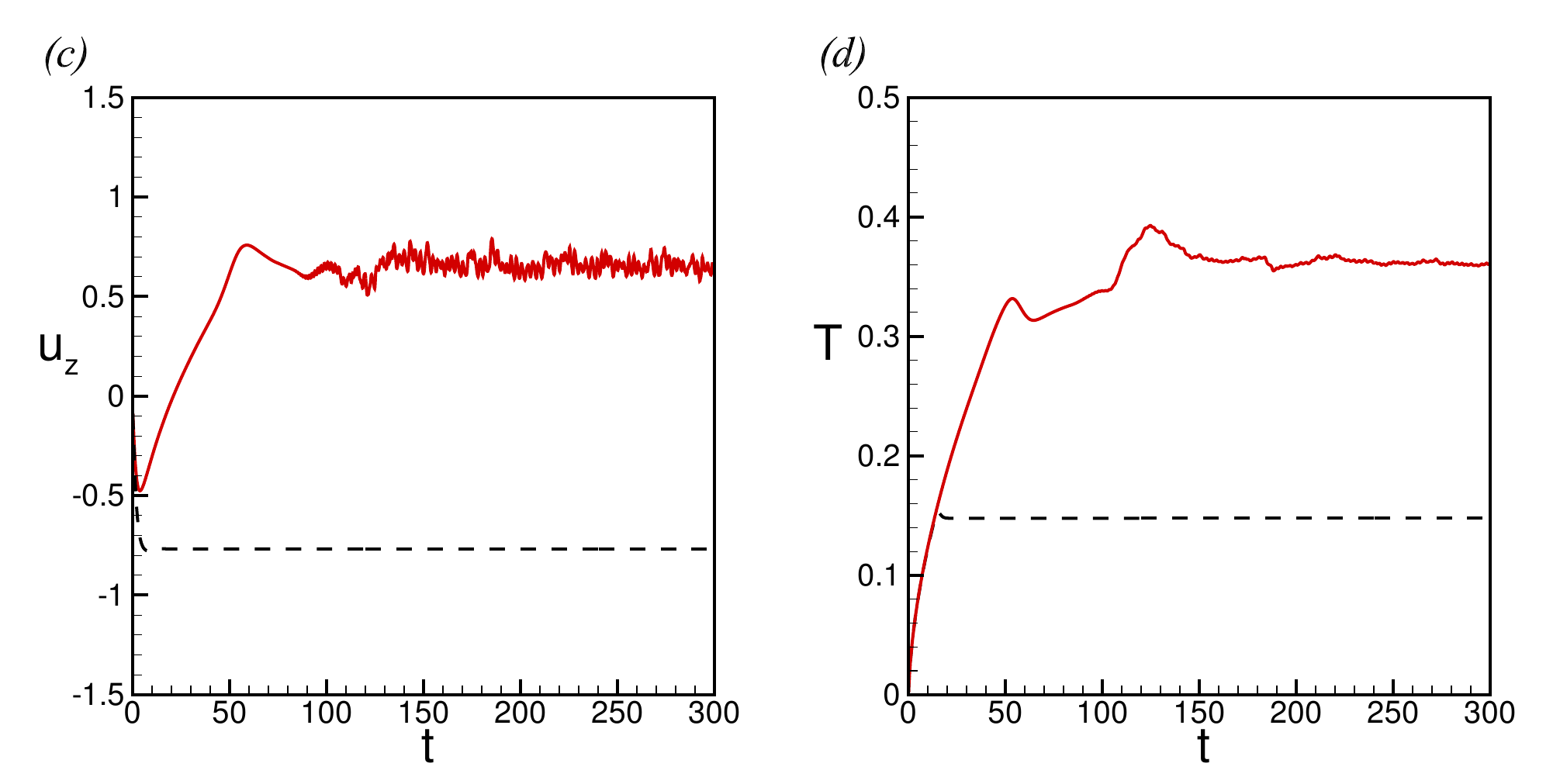}
\caption{Temporal evolution of stable and unstable flow regimes in Fig.~\ref{fig2}.  Average kinetic energy \emph{(a)}, mean temperature \emph{(b)}, and signals of vertical velocity $u_z$ \emph{(c)} and temperature $T$ \emph{(d)} at the point $y=-0.75$, $z=15$ are shown for $\Ha=10^4$, $\Rey=2\times10^4$, $\Gras=10^9$ (red solid lines), and $\Gras=10^8$ (black dashed lines).}
\label{fig3}
\end{center}
\end{figure*}

The temporal evolution of the flows in Fig.~\ref{fig2} is illustrated in Fig.~\ref{fig3} using the average kinetic energy (\ref{ke}), mean temperature (\ref{tm}), and the signals of $u_z$ and $T$ at the point $y=-0.75$, $z=15$. We see strong difference between the stable and unstable cases. At the fully developed flow stage, the latter shows much higher mean and point temperatures, strong upward velocity (see Fig.~\ref{fig3}c) and apparently irregular fluctuations of velocity and kinetic energy.

\begin{figure*}
\begin{center}
\includegraphics[width=0.48\textwidth]{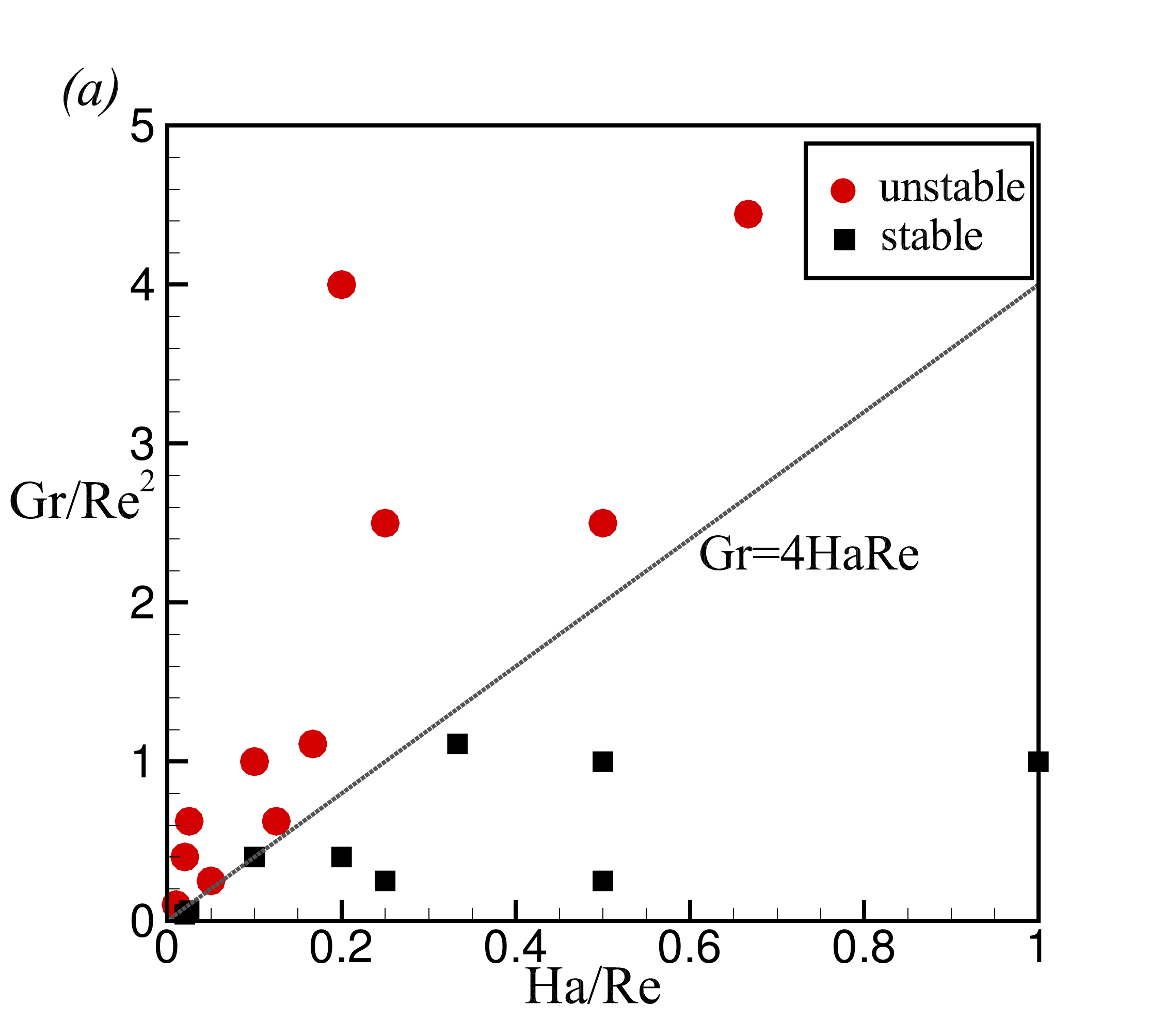}
\includegraphics[width=0.48\textwidth]{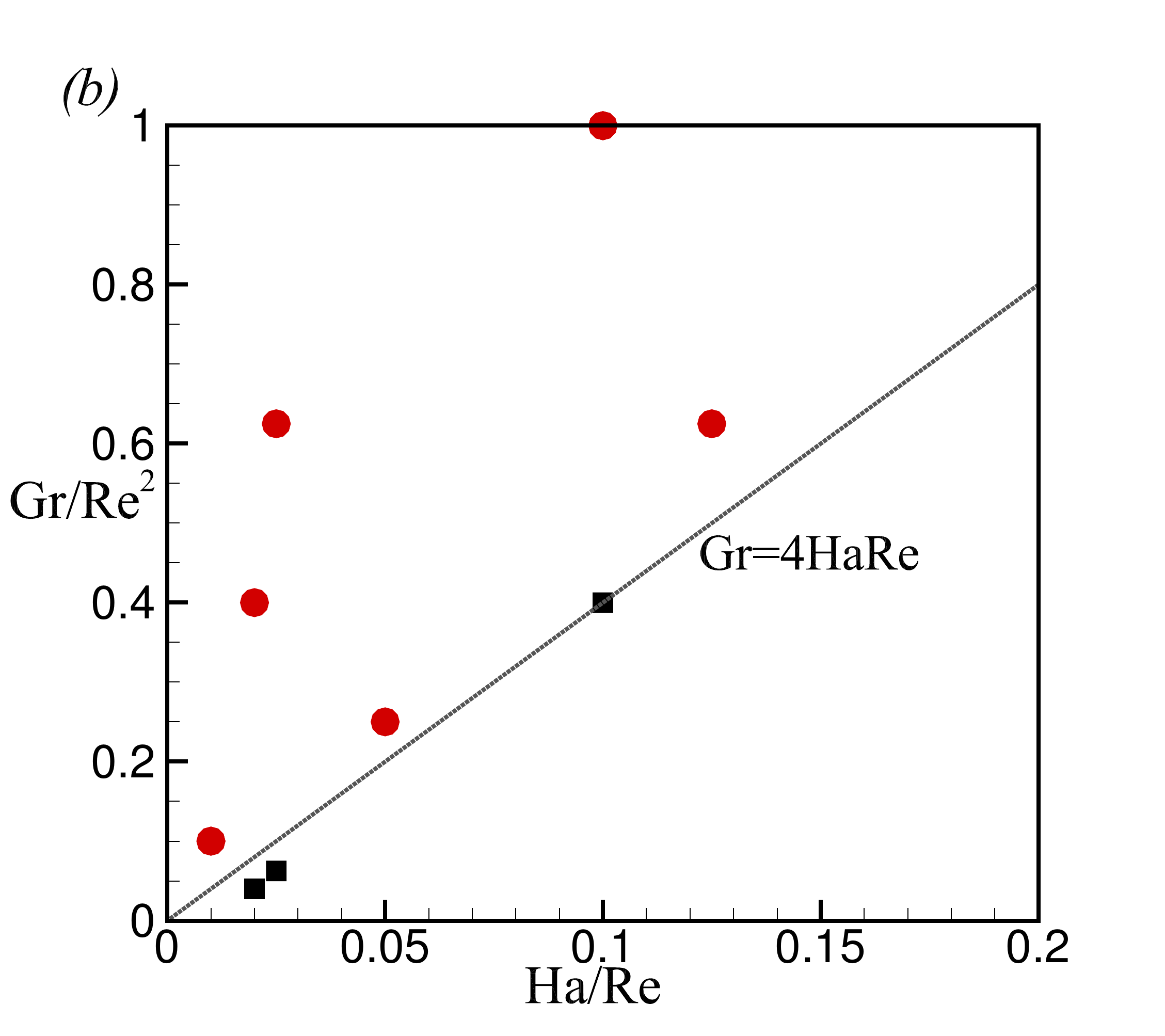}
\caption{Summary charts of the flow states for all the cases \emph{(a)} and cases at lower values of $\Ha/\Rey$ and $\Gras/\Rey^2$ \emph{(b)}.}
\label{fig4}
\end{center}
\end{figure*}

\begin{figure*}[!]
\begin{center}
\includegraphics[width=0.24\textwidth,trim=4 4 4 4,clip]{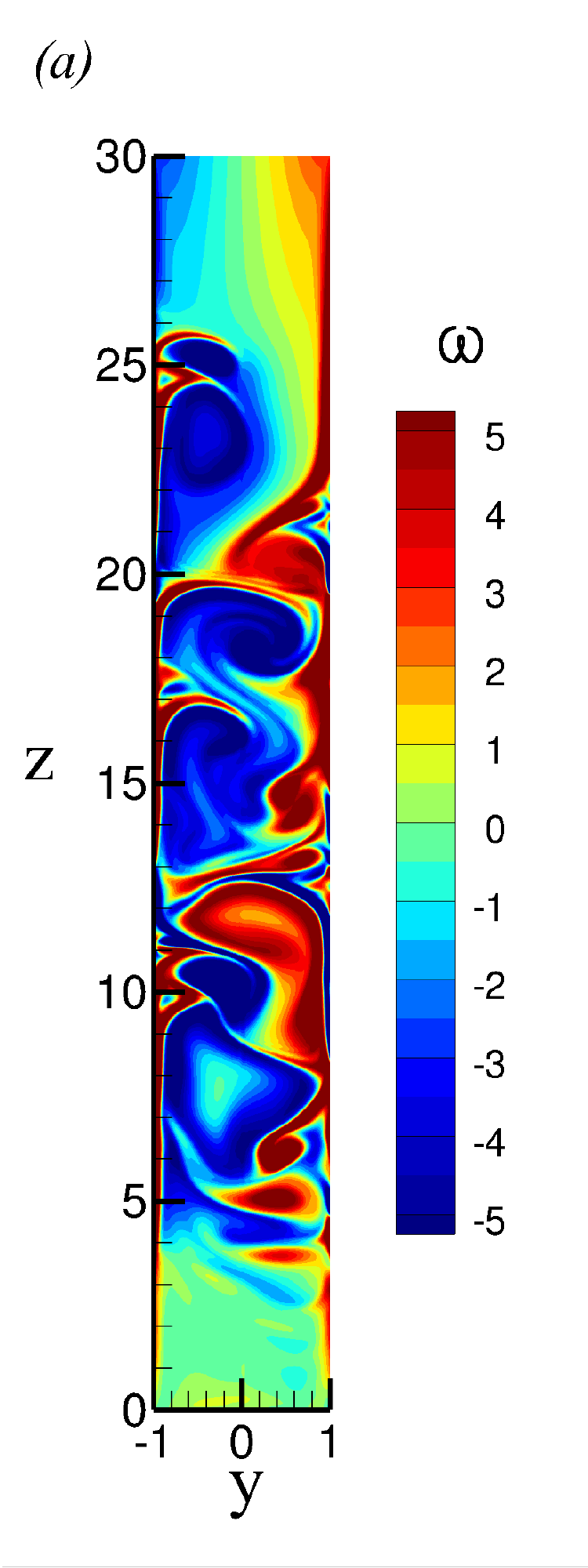}
\includegraphics[width=0.24\textwidth,trim=4 4 4 4,clip]{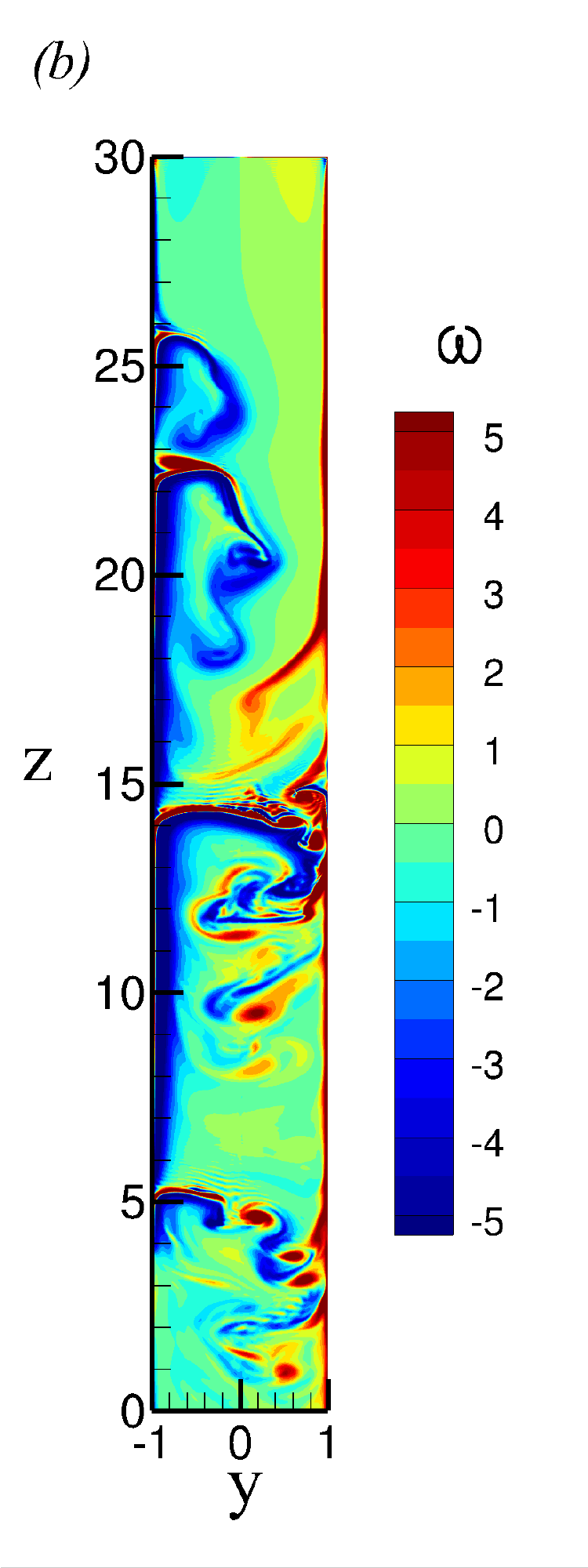}
\includegraphics[width=0.24\textwidth,trim=4 4 4 4,clip]{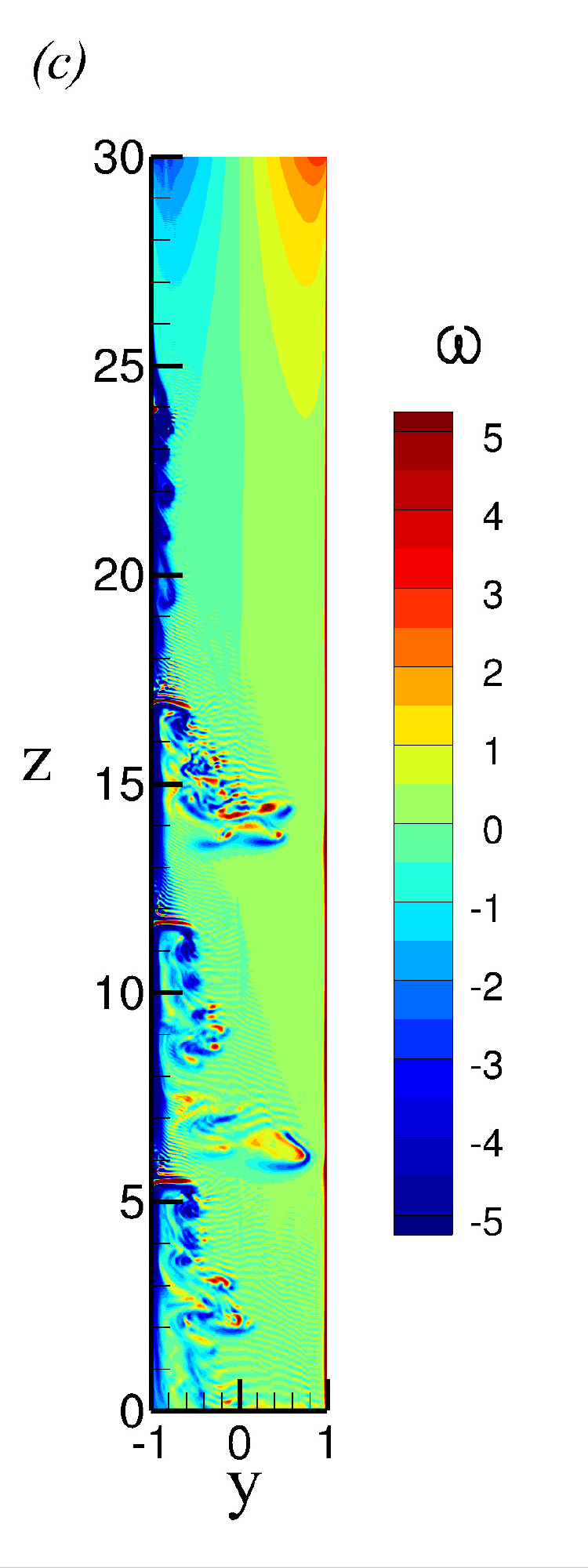}
\caption{Instantaneous distributions of vorticity $\omega$ in the flows at $\Gras=10^6$, $\Ha=100$, $\Rey=500$ \emph{(a)}, $\Gras=10^8$, $\Ha=1000$, $\Rey=5000$ \emph{(b)} and $\Gras=10^{10}$, $\Ha=10^4$, $\Rey=5\times10^{4}$ \emph{(c)}.}
\label{fig5}
\end{center}
\end{figure*}

\begin{figure*}[!]
\begin{center}
\includegraphics[width=0.7\textwidth,trim=4 4 4 4,clip]{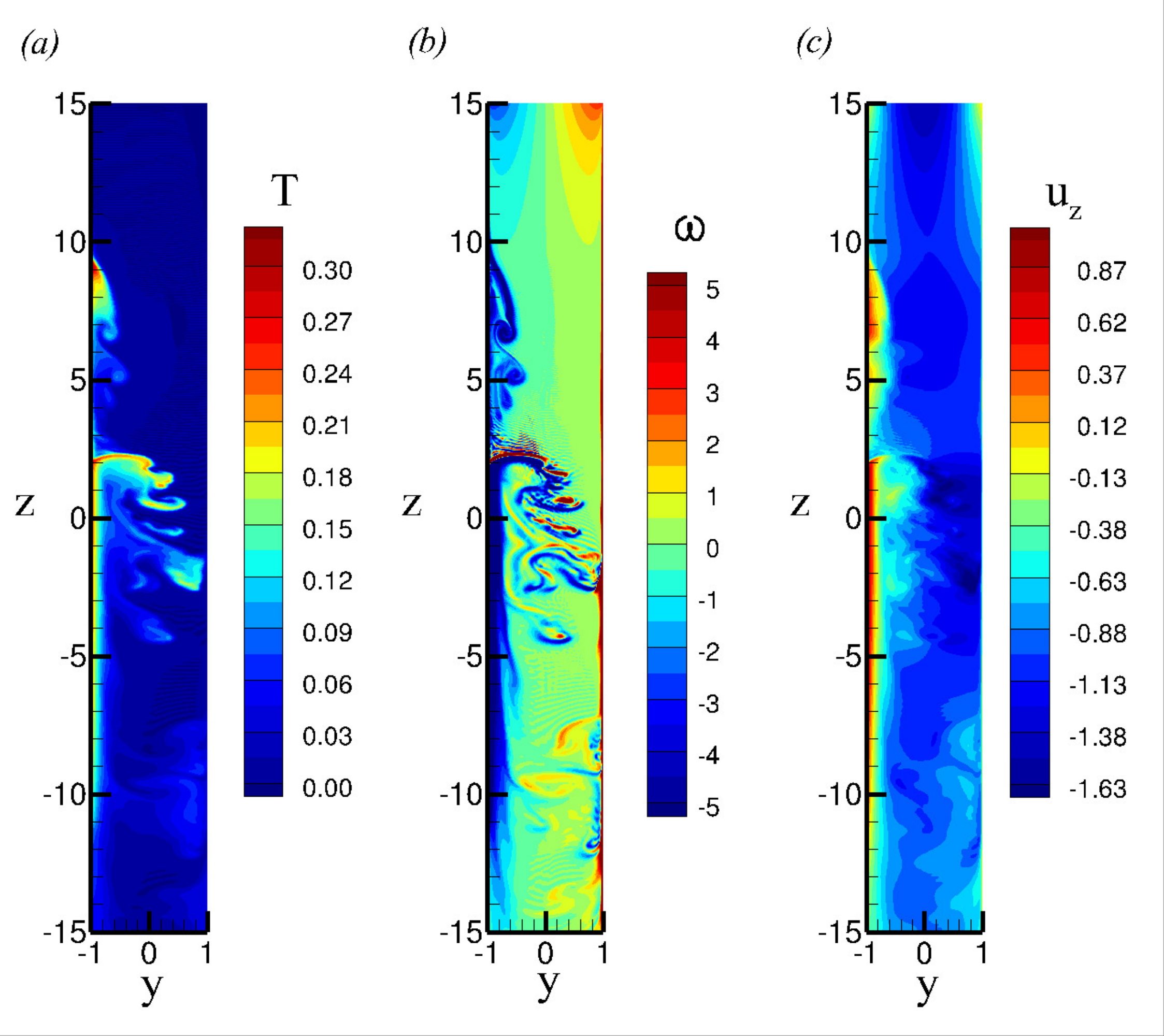}\\
\includegraphics[width=0.45\textwidth]{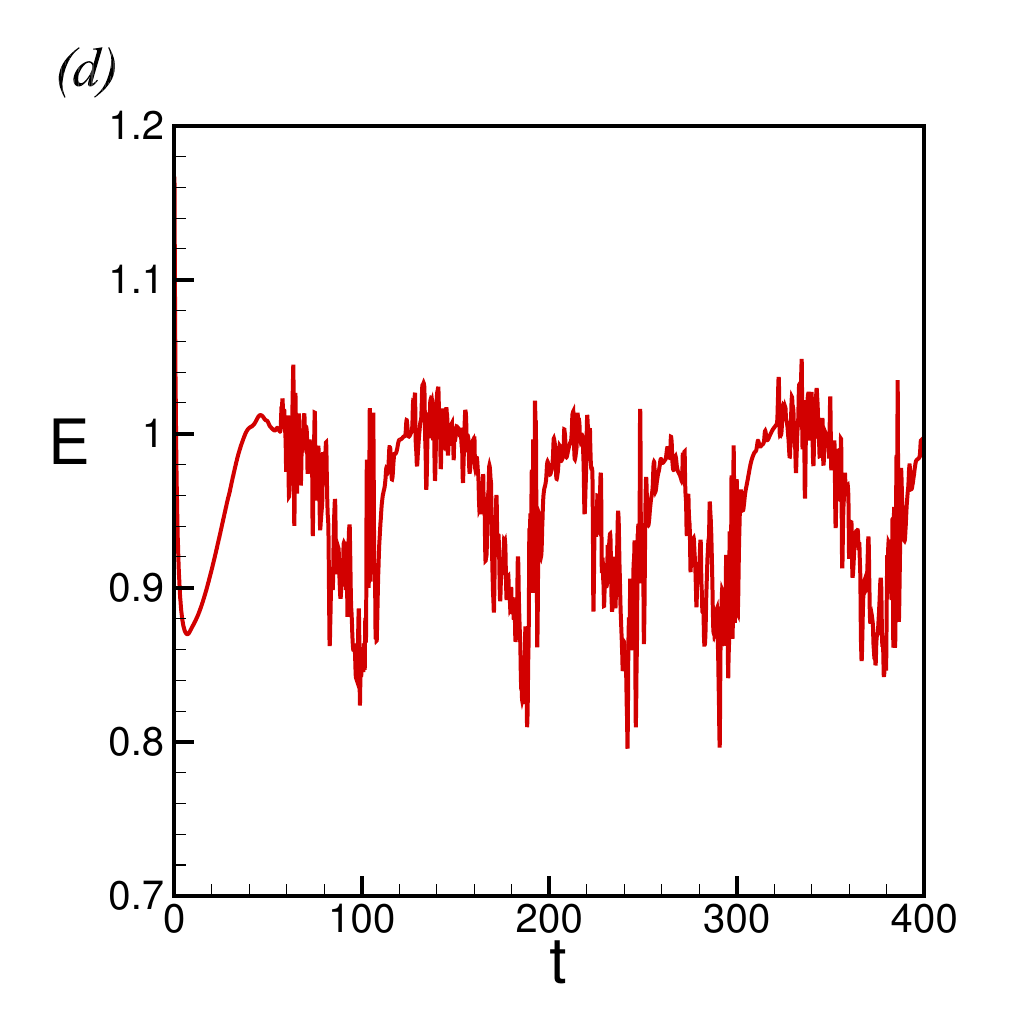}
\includegraphics[width=0.45\textwidth]{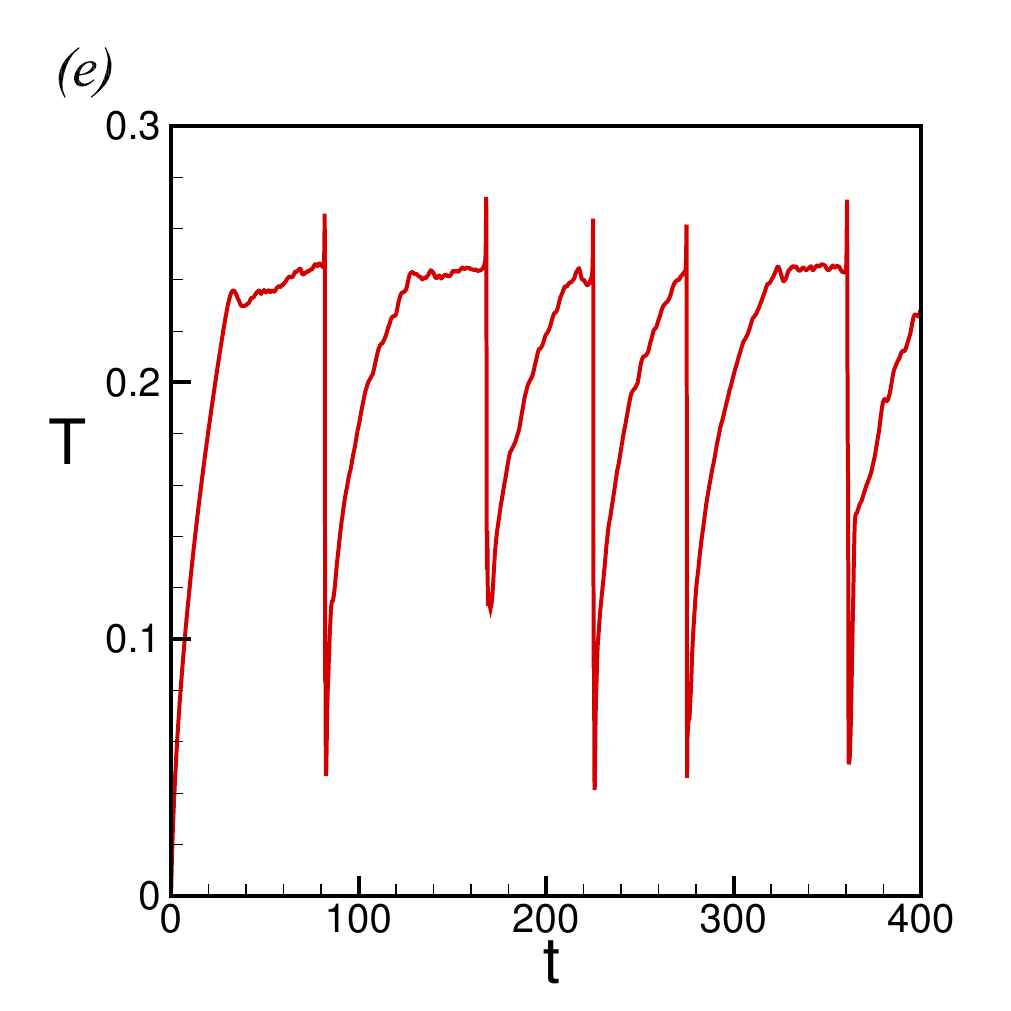}
\caption{Instantaneous distributions of temperature $T$ \emph{(a)}, vorticity $\omega$ \emph{(b)}, amplitude of vertical velocity $u_z$ \emph{(c)}, signal of average kinetic energy \emph{(d)} and point signal of temperature $T$ at $y=-0.75$, $z=15$ \emph{(e)} in the flow at $\Ha=5000$, $\Rey=2\times10^4$, $\Gras=10^9$.} 
\label{fig6}
\end{center}
\end{figure*}

\begin{figure*}[!]
\begin{center}
\includegraphics[width=0.7\textwidth,trim=4 4 4 4,clip]{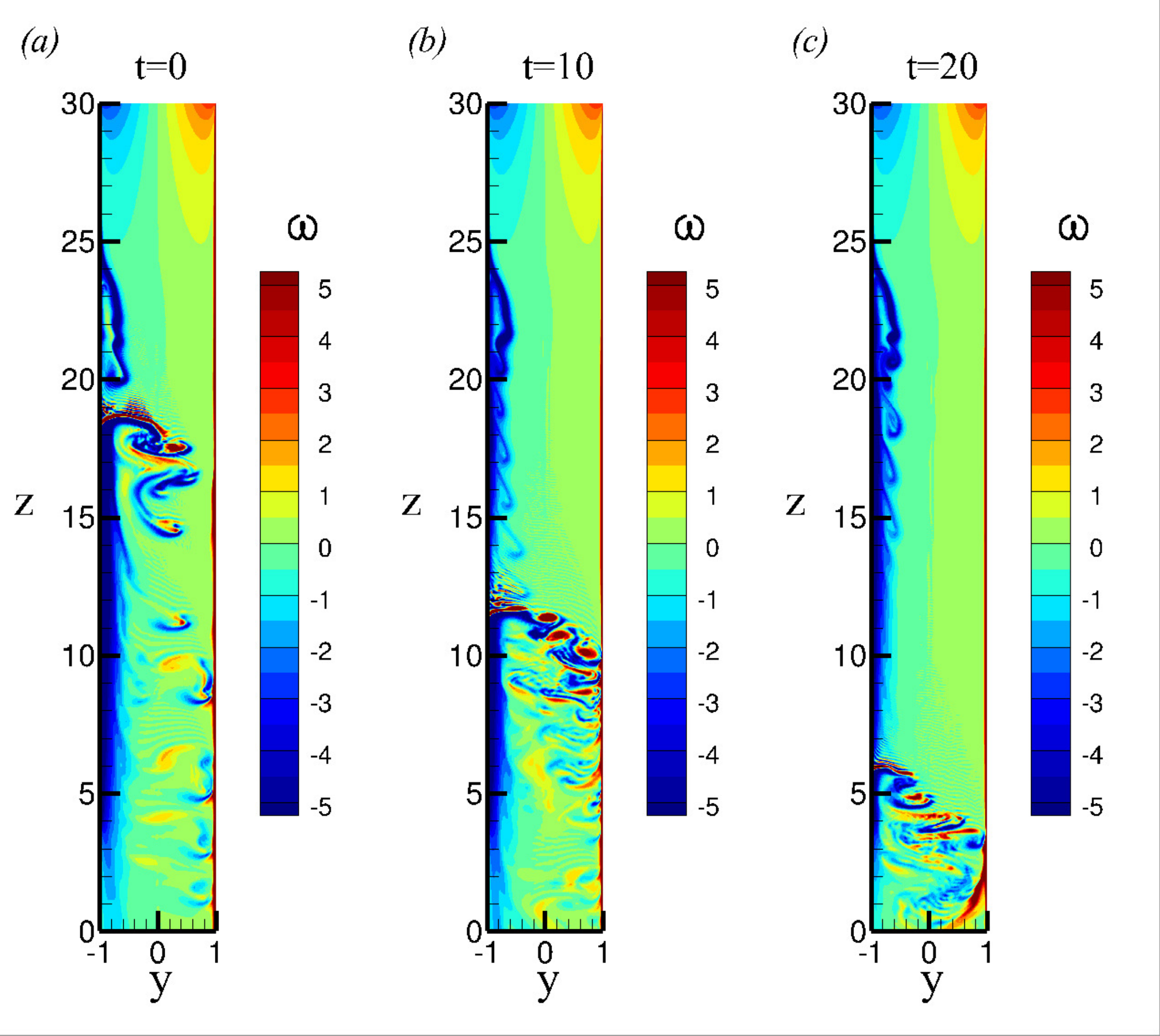}
\caption{Evolution of vorticity $\omega$ in the flow at $\Gras=10^9$, $\Ha=5000$, $\Rey=2\times10^4$. Non-dimensional time at which snapshot was taken is shown above the figure.} 
\label{fig7}
\end{center}
\end{figure*}

\begin{figure*}[!]
\begin{center}
\includegraphics[width=0.7\textwidth,trim=4 4 4 4,clip]{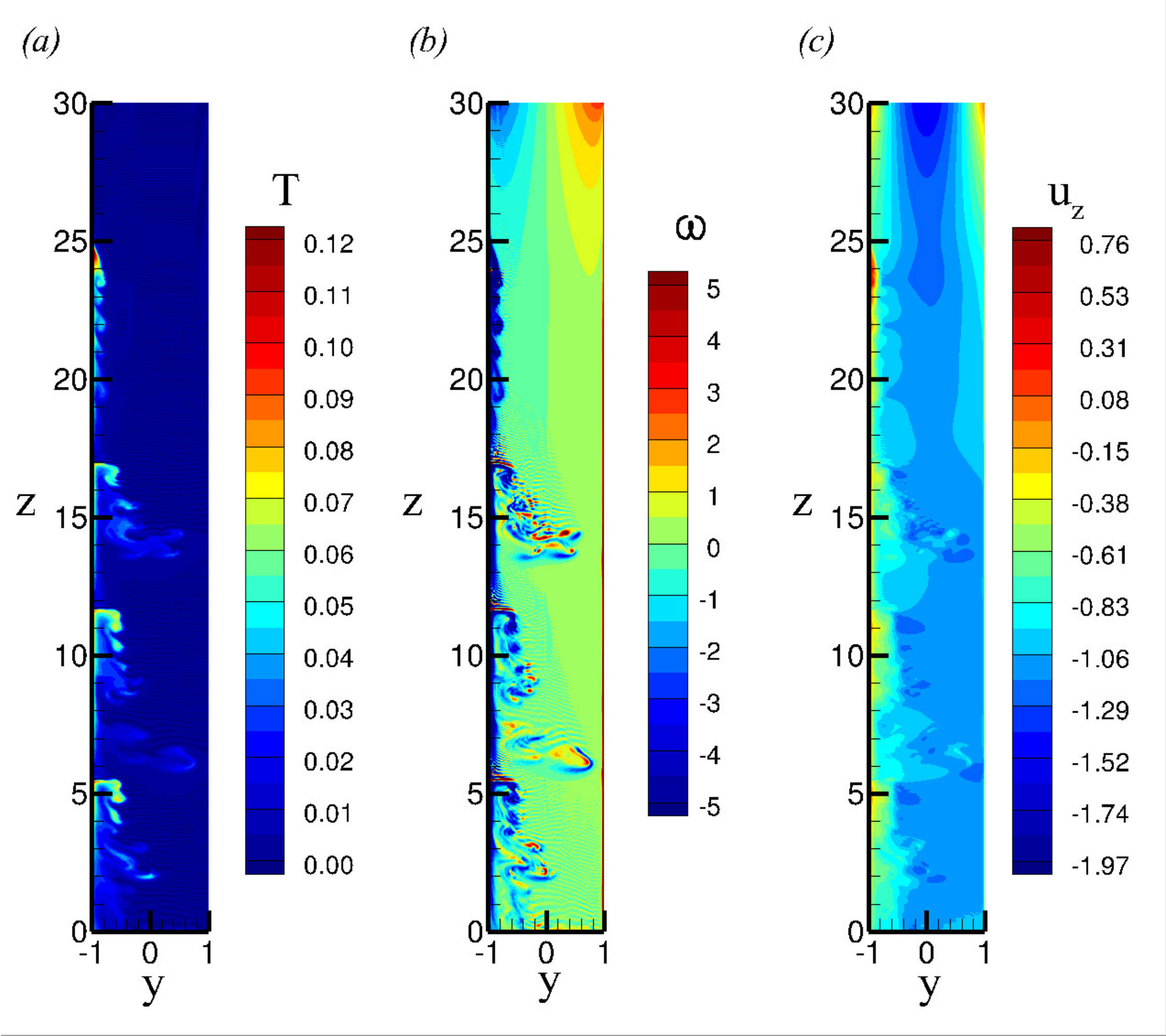}\\
\includegraphics[width=0.45\textwidth]{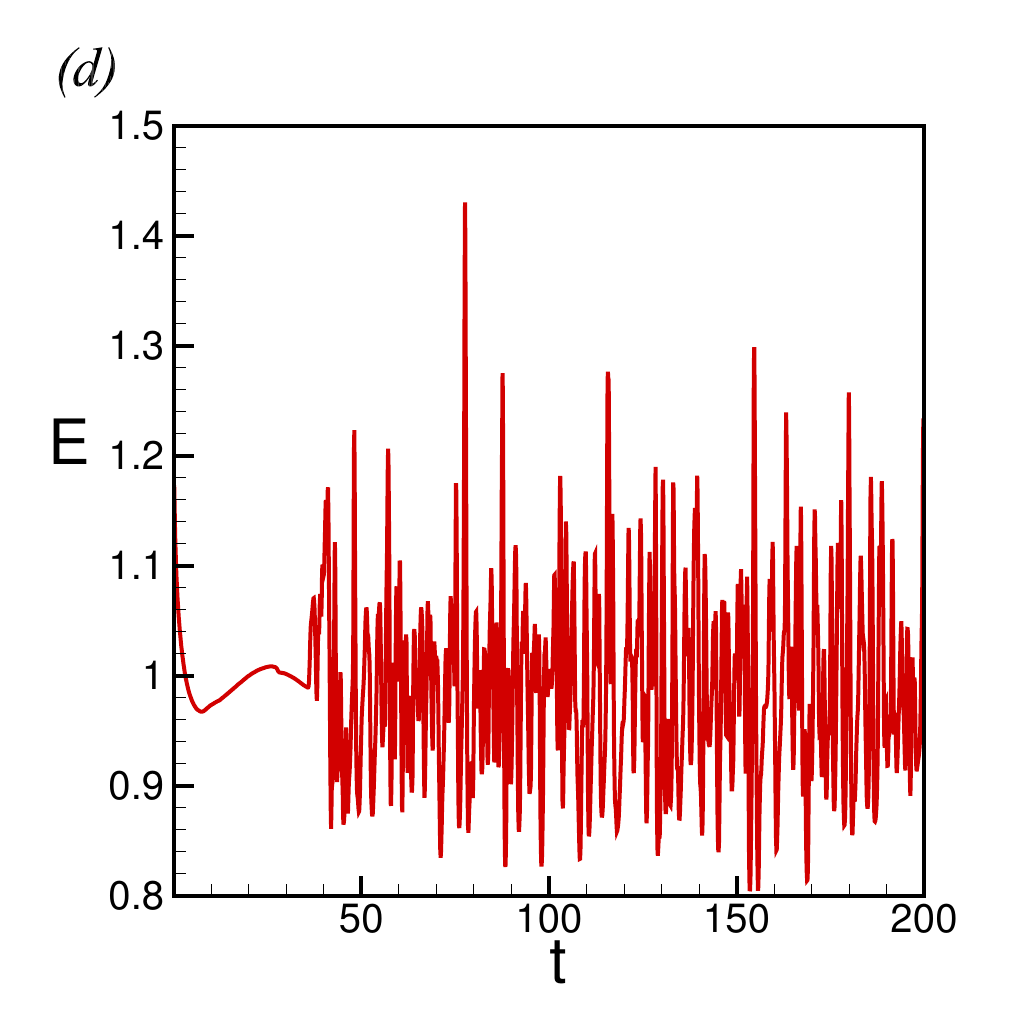}
\includegraphics[width=0.45\textwidth]{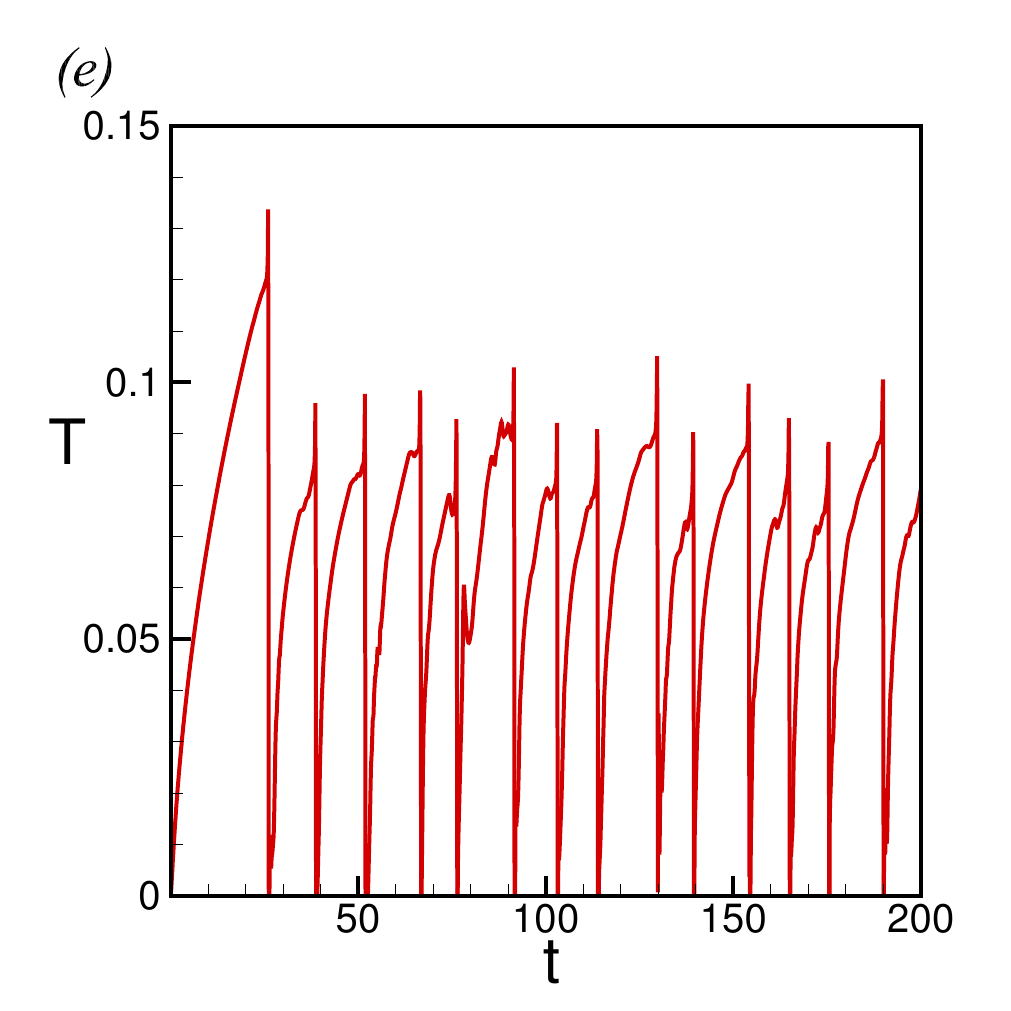}
\caption{Instantaneous distributions of temperature $T$ \emph{(a)}, vorticity $\omega$ \emph{(b)}, amplitude of vertical velocity $u_z$ \emph{(c)}, signal of average kinetic energy \emph{(d)} and point signal of temperature $T$ at $y=-0.75$, $z=15$ \emph{(e)} in the flow at $\Ha=10^4$, $\Rey=5\times10^4$, $\Gras=10^{10}$.} 
\label{fig8}
\end{center}
\end{figure*}

\begin{figure*}[!]
\begin{center}
\includegraphics[width=0.7\textwidth,trim=4 4 4 4,clip]{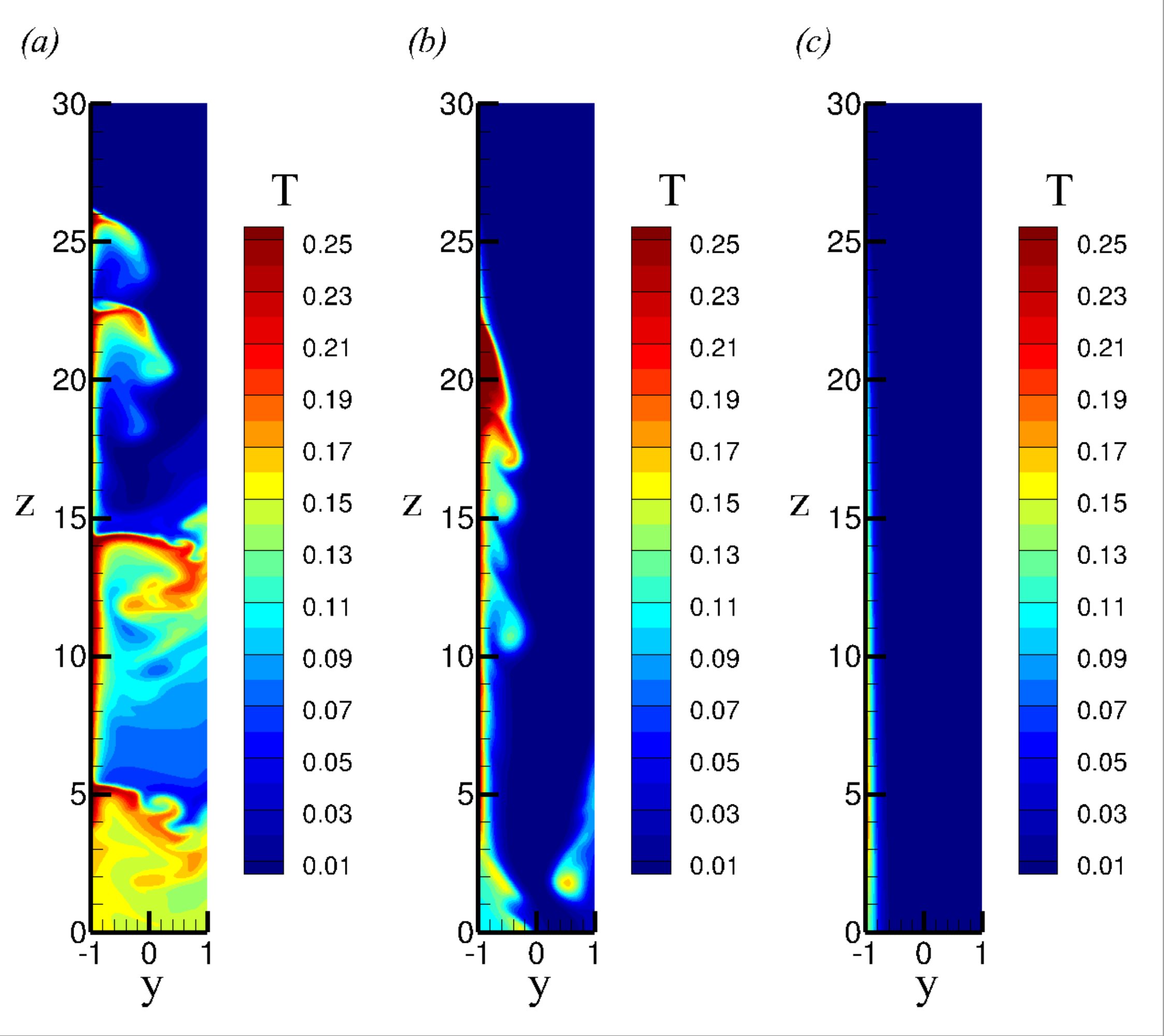}
\caption{Instantaneous distributions of temperature $T$ in the flows $\Gras=10^8$, $\Ha=1000$, $\Rey=5000$ \emph{(a)}, $\Rey=2\times10^4$ \emph{(b)} and $\Rey=4\times10^4$ \emph{(c)}.}
\label{fig9}
\end{center}
\end{figure*}

\begin{figure*}
\begin{center}
\includegraphics[width=0.9\textwidth]{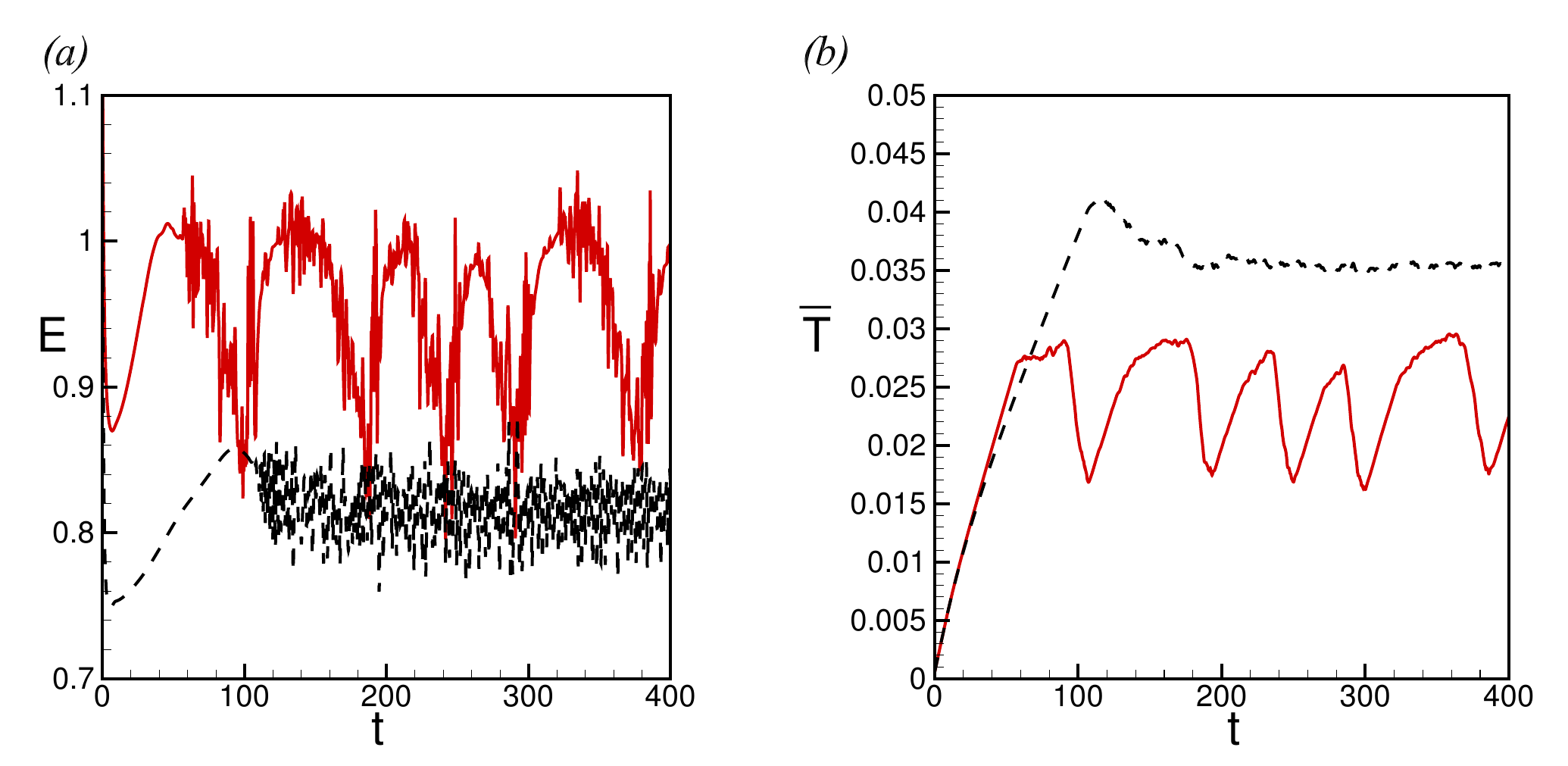}
\caption{Average kinetic energy \emph{(a)} and mean temperature \emph{(b)} in the flow at $\Gras=10^9$, $\Rey=5000$, $\Ha=5000$ (red, solid lines) and $\Ha=10^4$ (black, dashed lines).}
\label{fig10}
\end{center}
\end{figure*}

\begin{figure*}[!]
\begin{center}
\includegraphics[width=0.24\textwidth,trim=4 4 4 4,clip]{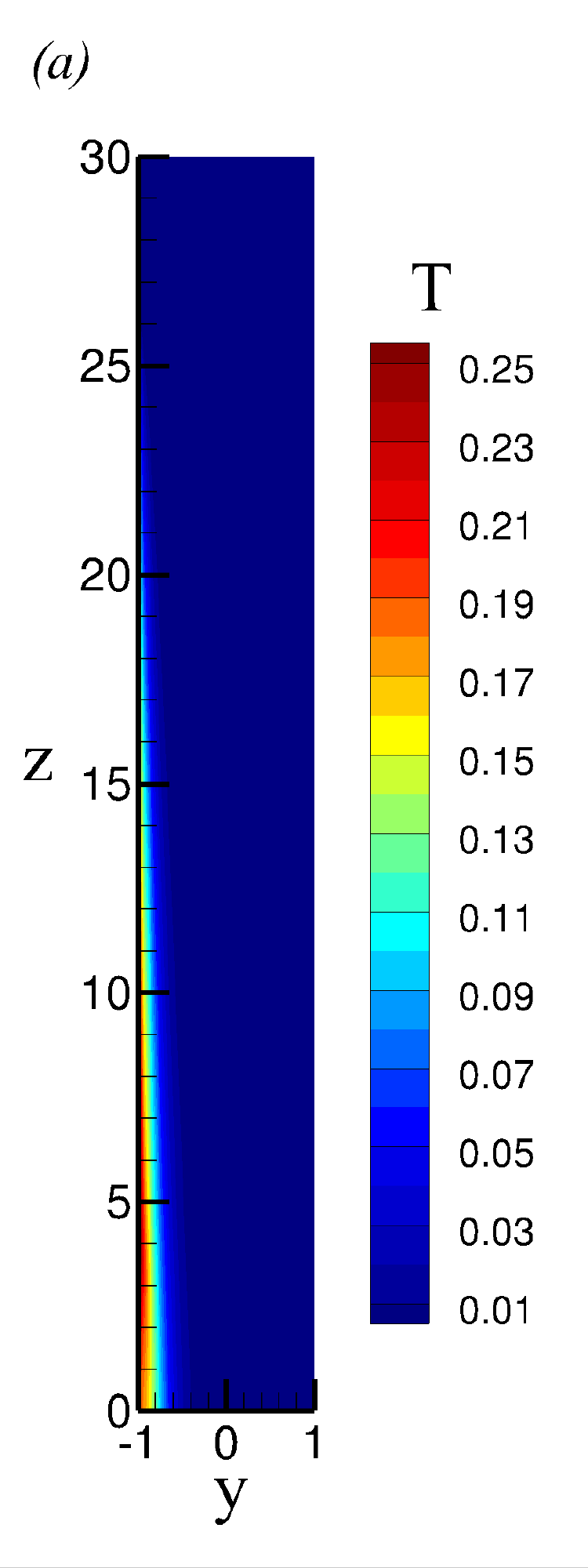}
\includegraphics[width=0.24\textwidth,trim=4 4 4 4,clip]{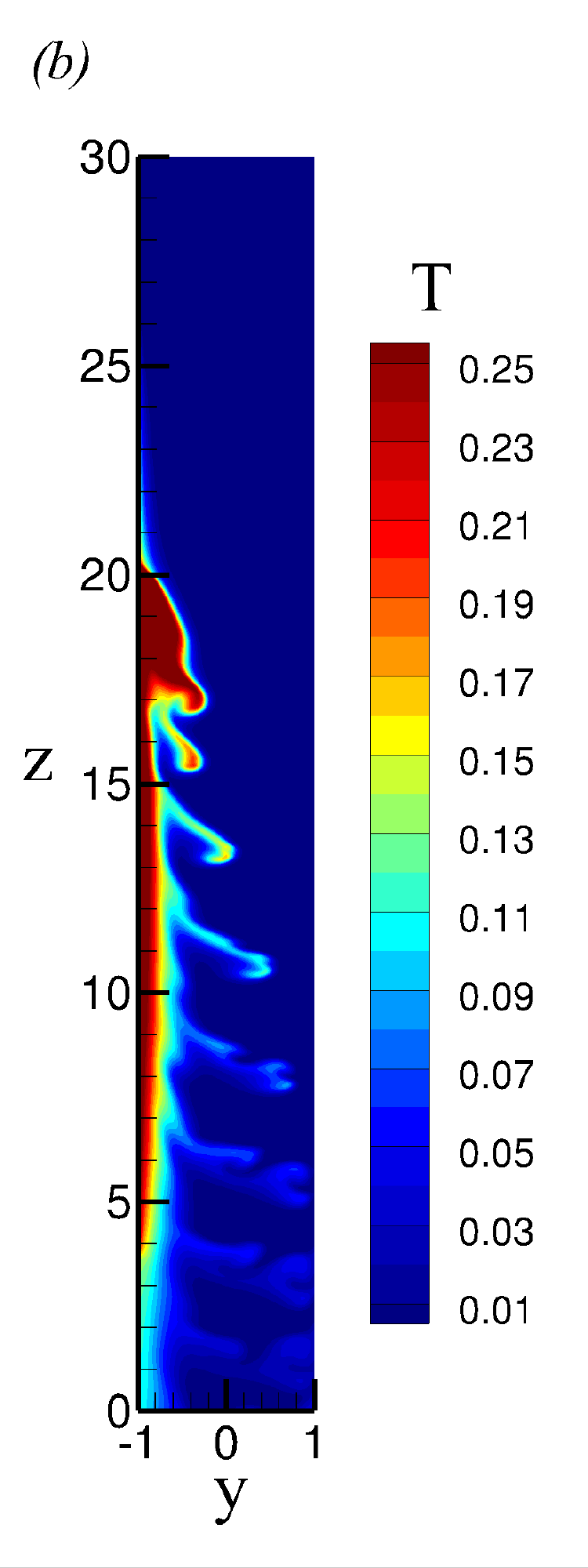}
\includegraphics[width=0.24\textwidth,trim=4 4 4 4,clip]{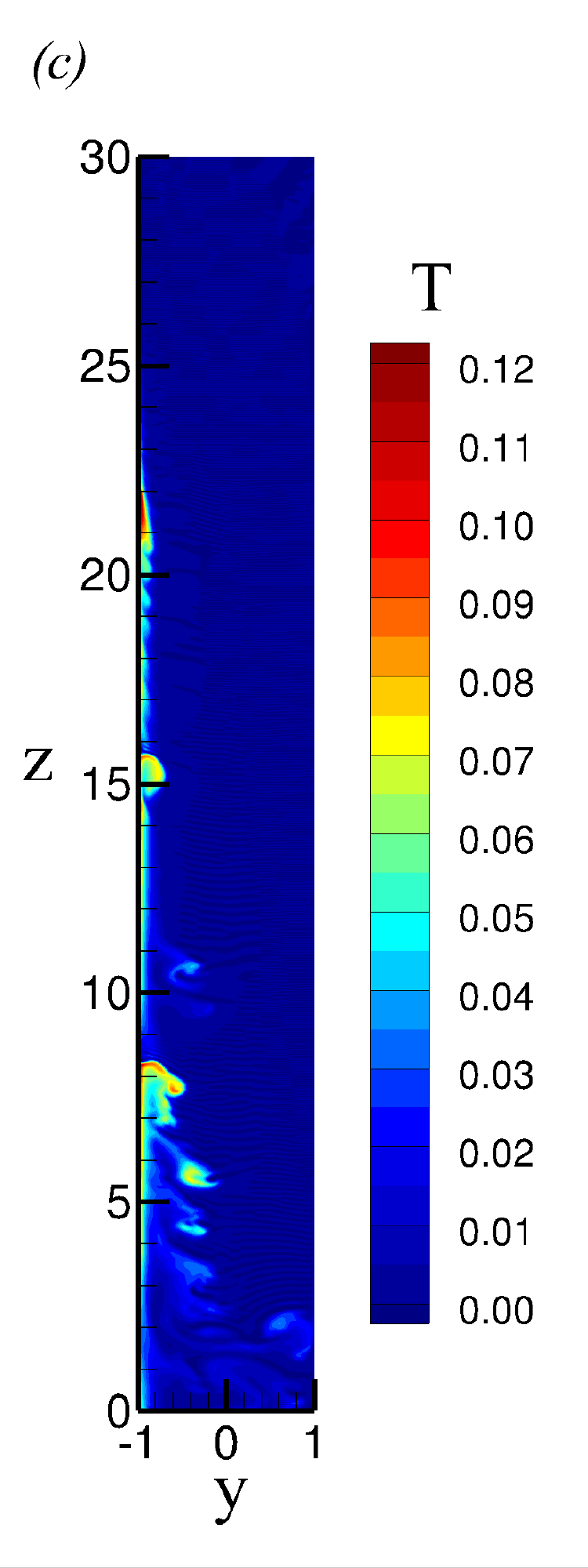}
\includegraphics[width=0.24\textwidth,trim=4 4 4 4,clip]{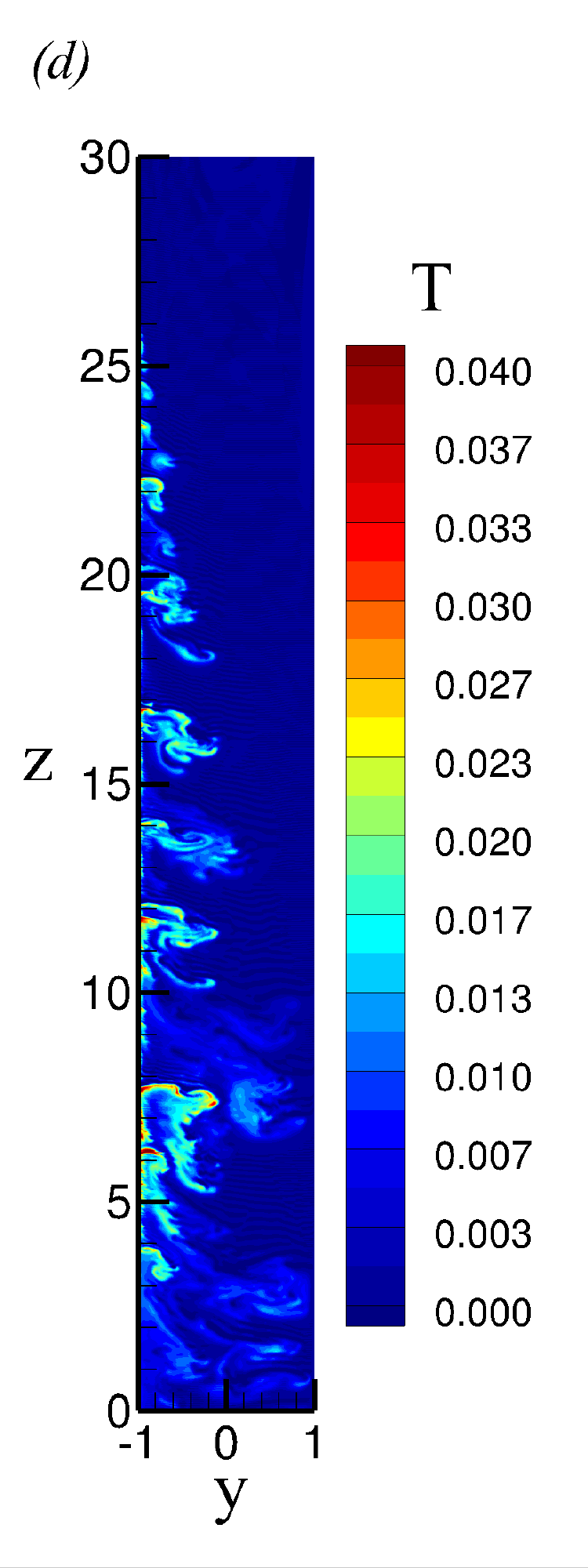}
\caption{Instantaneous distributions of temperature $T$ in the flows at $\Ha=10^4$, $\Rey=2\times10^4$, $\Gras=10^8$ \emph{(a)}, $\Gras=10^9$ \emph{(b)}, and at $\Ha=10^4$, $\Rey=5\times10^4$, $\Gras=10^{10}$ \emph{(c)}, $\Gras=10^{11}$ \emph{(d)}. Note that the isolevels of temperature are different at different $\Gras$.}
\label{fig11}
\end{center}
\end{figure*}

\begin{figure*}[!]
\begin{center}
\includegraphics[width=0.7\textwidth,trim=4 4 4 4,clip]{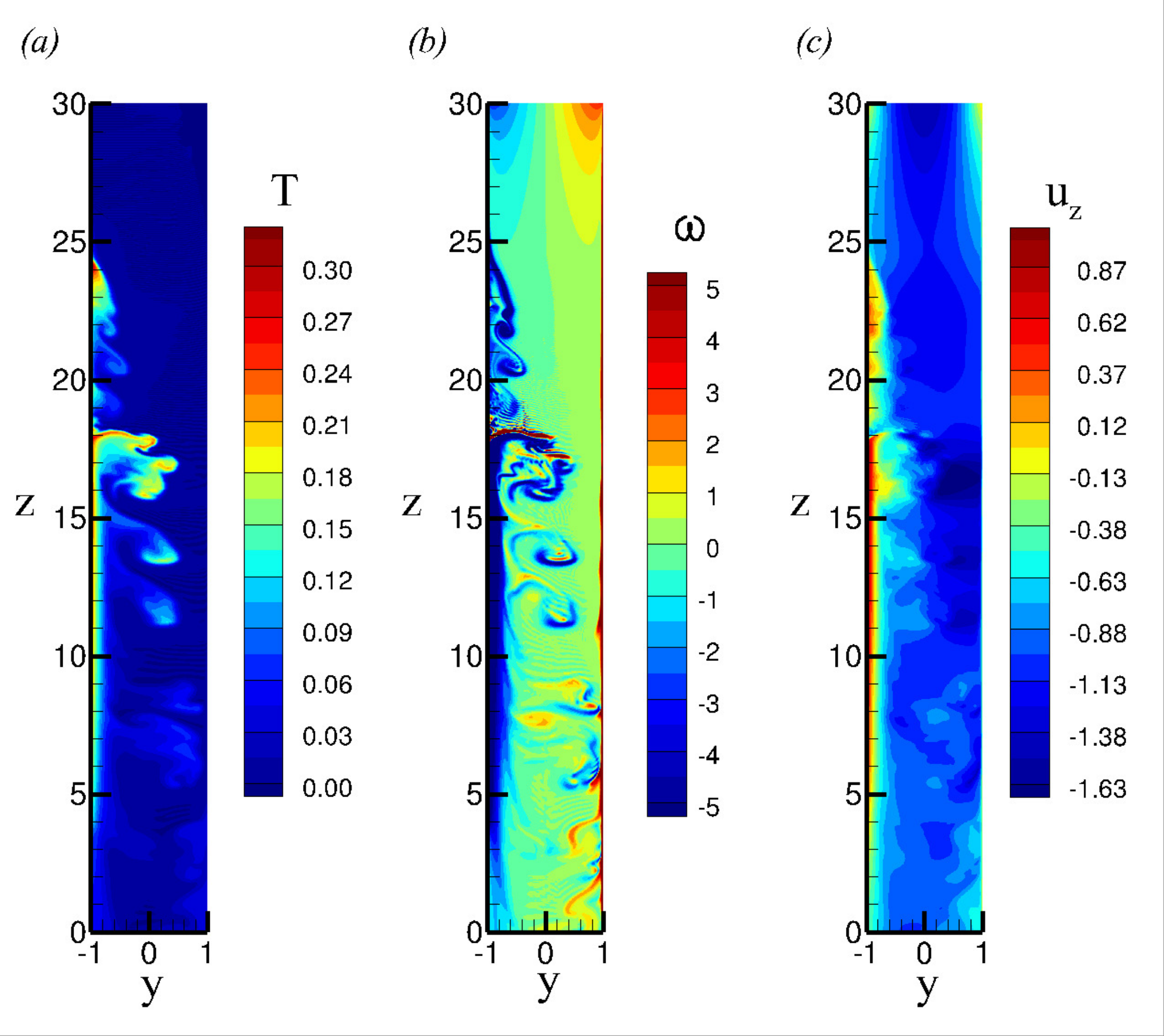}\\
\includegraphics[width=0.7\textwidth,trim=4 4 4 4,clip]{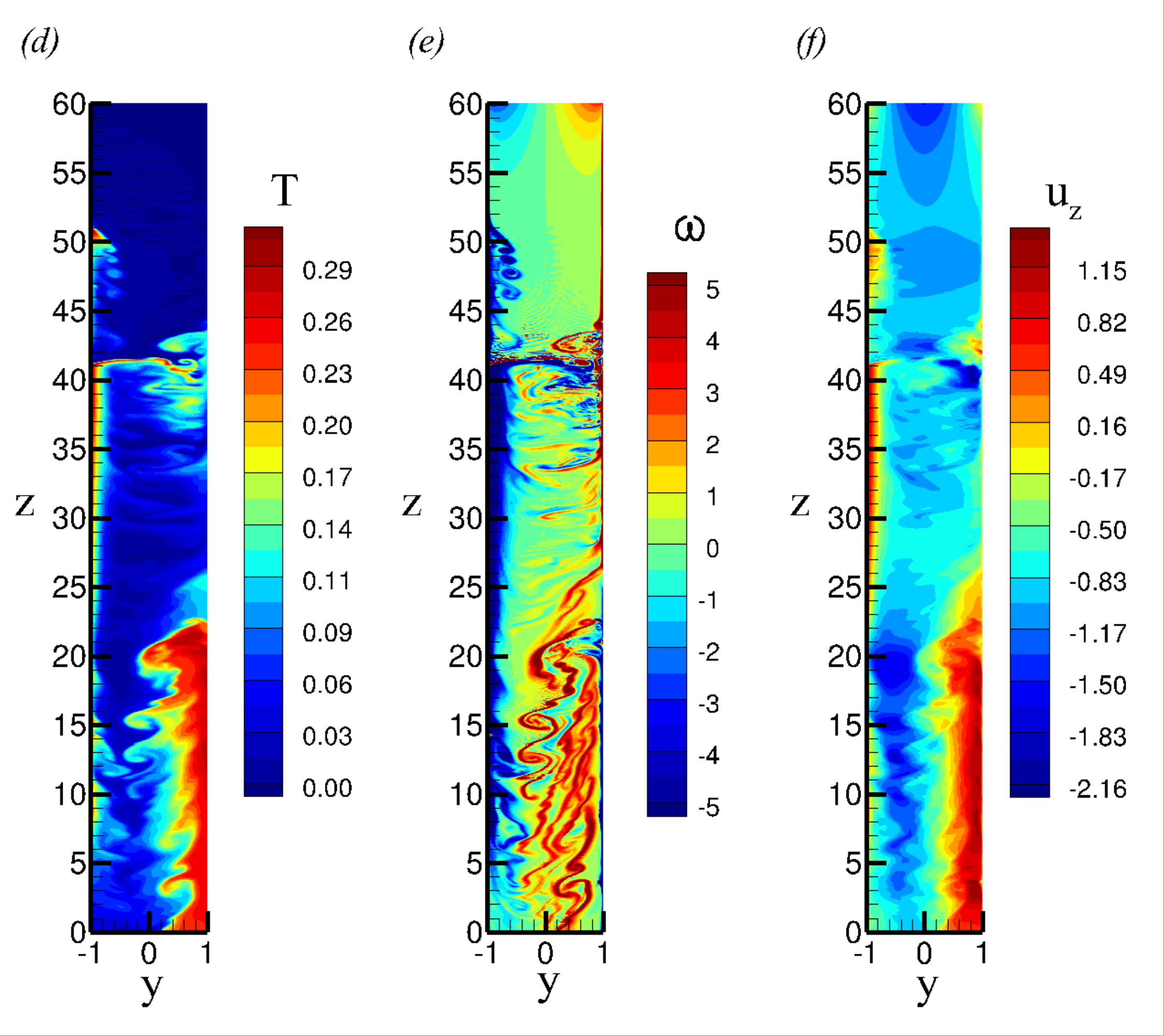}
\caption{Instantaneous distributions of temperature $T$, vorticity $\omega$ and amplitude of vertical velocity $u_z$ in the flows at $\Ha=5000$, $\Rey=2\times10^4$, $\Gras=10^9$, $L_z=30$ \emph{(a)-(c)} and $L_z=60$ \emph{(d)-(f)}.} 
\label{fig12}
\end{center}
\end{figure*}

\begin{figure*}
\begin{center}
\includegraphics[width=0.45\textwidth]{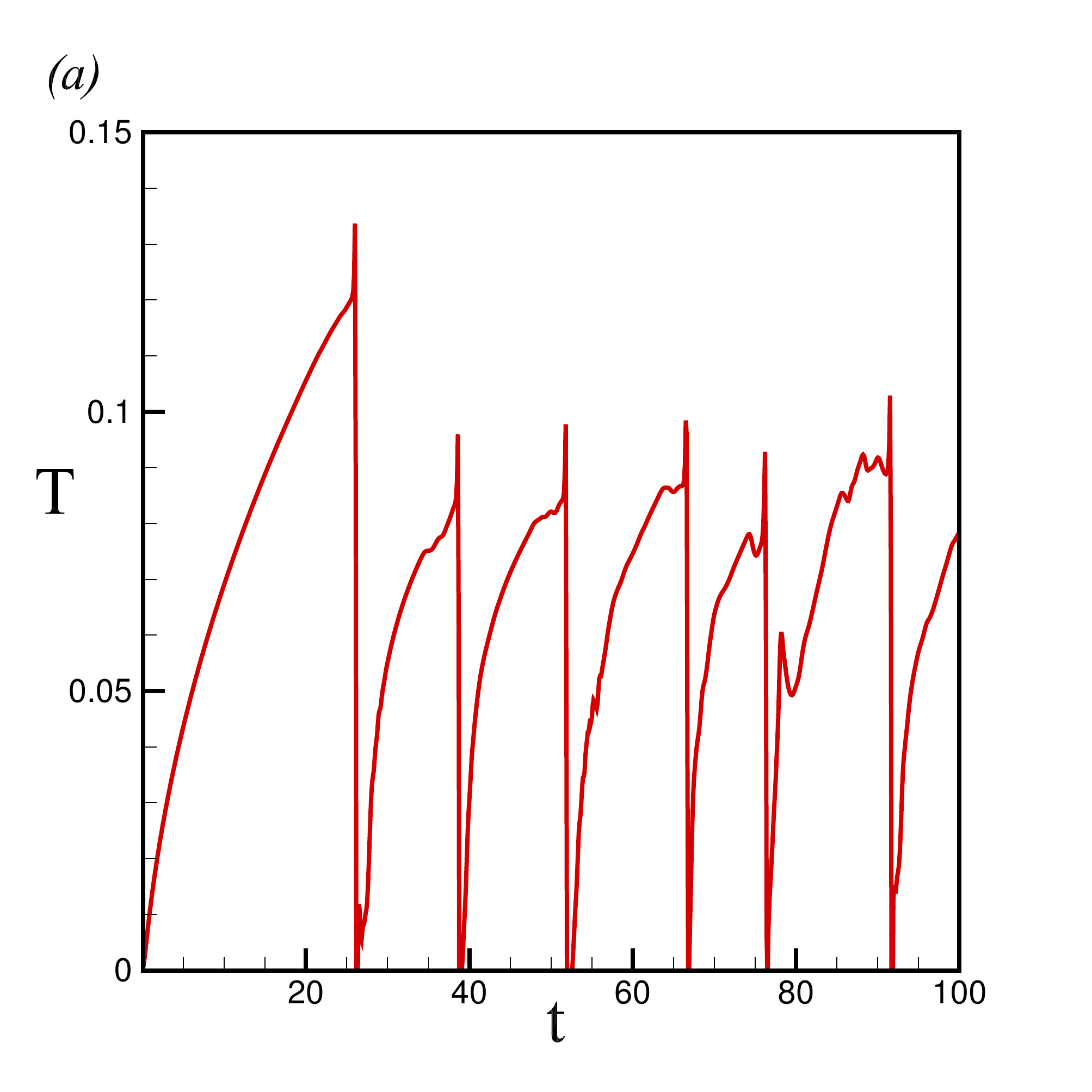}
\includegraphics[width=0.45\textwidth]{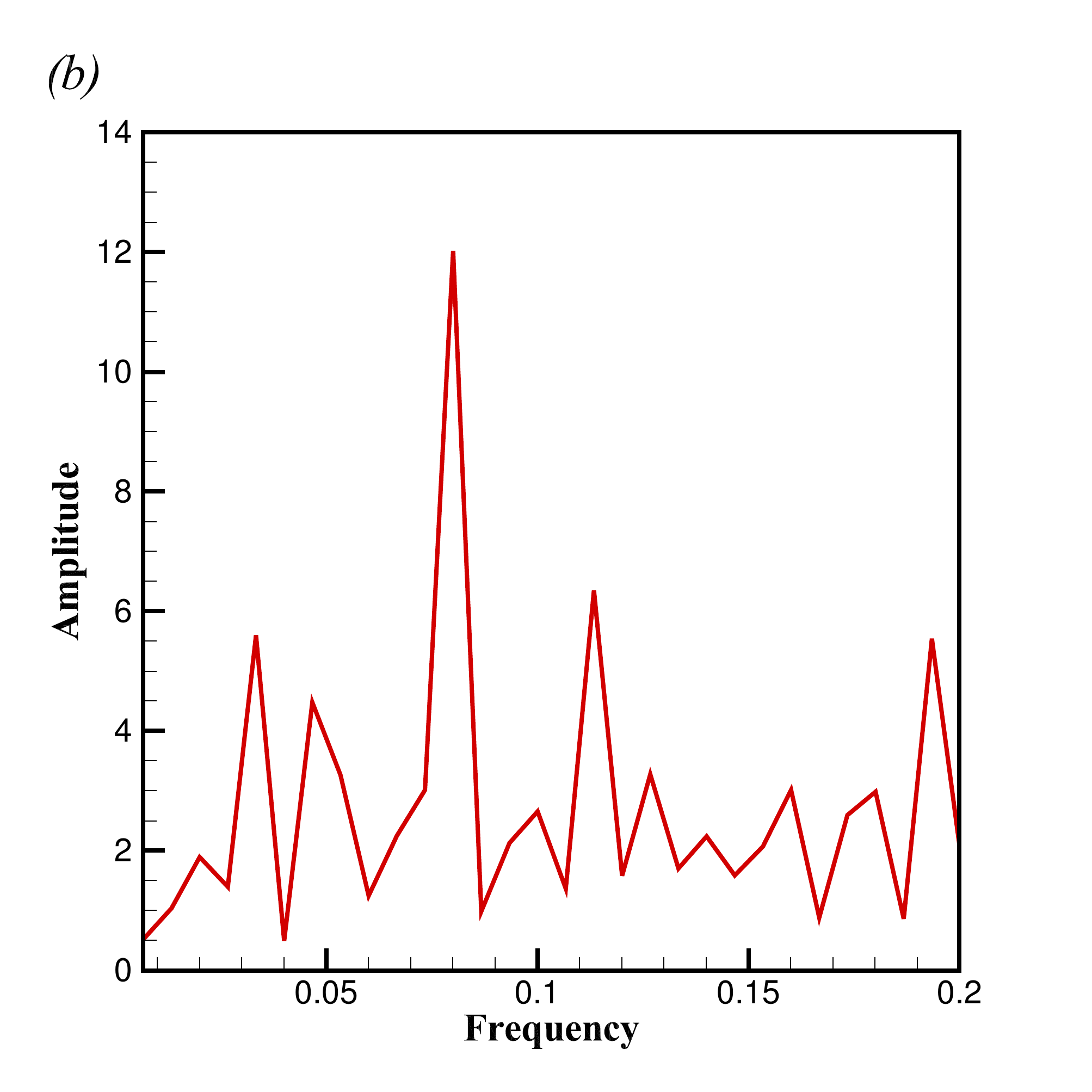}
\caption{Point signal of temperature $T$ at $y=-0.75$, $z=15$\emph{(a)} and the spectral decomposition \emph{(b)} at $\Gras=10^{10}$, $\Ha=10^4$ and $\Rey=5\times10^4$.}
\label{fig13}
\end{center}
\end{figure*}

We now move to detailed discussion of the various flow patterns observed in our study. 
The performed simulations are summarized in Fig.~\ref{fig4} and table \ref{tab2}.  The most extensive studies are performed at $\Gras=10^8$ and $\Gras=10^9$. 
The  flow structure and time evolution at different values of $\Gras$, $\Ha$ and $\Rey$ are illustrated in Figs.~\ref{fig5}-\ref{fig12}. 
The duct of length $L_z=30$ is primarily studied, since it is, as we discuss below, particularly relevant to the applications to the liquid mental blanket. The effect of duct length is discussed at the end of this section and illustrated Fig.~\ref{fig13}.

In order to better quantify the jets and the secondary roll structures, we show in Table \ref{tab2}, in addition to $E$ and $\overline{T}$, the kinetic energy of the transverse velocity component and the maximum difference in the vertical velocity:
\begin{equation}\label{trans_ke}  
 E_y  = \frac{1}{A}\int_{0}^{L}\int_{-1}^1 u_y^2 dydz,
\end{equation}
\begin{equation}\label{uzmax} 
{\Delta u_z}^{max} = {u_z}^{up} - {u_z}^{down}.
\end{equation}
Here ${u_z}^{up}$ and ${u_z}^{down}$ are the maximum vertical velocities in the upward and downward parts of the flow in the fully developed steady state (for stable flows) or right before the jet breakdown (for unstable flows). As an example, the flow in Figs.~\ref{fig2}d-f has jets with ${\Delta u_z}^{max}=2.56$ and the instability rolls with $E_y=5.74\times 10^{-3}$. Comparing the latter with the total kinetic energy of the flow $E=0.82$, we find that the fully developed rolls are quite weak in comparison with the mean flow.

 \tabcolsep=0.3cm
\begin{table*}
\newcommand{\tabincell}[2]{\begin{tabular}{@{}#1@{}}#2\end{tabular}}
  \begin{center}
  \scalebox{0.8}{
  \begin{tabular}{@{}ccc|ccc|ccccc@{}}
  $\Gras$ & $\Ha$ & $\Rey$ & $\frac{\Gras}{\Rey^2}$ & $\frac{\Ha}{\Rey}$ & $\N$ & $E$ & $\overline{T}$ & ${\Delta u_z}^{max}$ & $E_y$ & \tabincell{c}{Flow\\type} \\ 
  \hline
  $10^6$ & 100 & 500 & 4.00  & 0.20 & 20 & 2.45 & $5.12\times10^{-1}$ & 6.11 & $4.80\times10^{-1}$&unstable \\
  $10^6$ & 100 & 1000 & 1.00  & 0.10 & 10 & 0.79 & $4.11\times10^{-1}$ & 6.00 & $1.09\times10^{-1}$&unstable \\
  \hline 
  $10^8$ & 1000 & 5000 & 4.00  & 0.20 & 200 & 1.22 & $6.09\times10^{-2}$ & 5.04 & $6.44\times10^{-2}$&unstable \\
  $10^8$ & 1000 & $10^4$ & 1.00  & 0.10 & 100 & 0.44 & $4.83\times10^{-2}$ & 3.28 & $1.70\times10^{-2}$&unstable \\
  $10^8$ & 1000 & $2\times10^4$ & 0.25  & 0.05 & 50 & 0.45 & $3.08\times10^{-2}$ & 2.22 & $4.22\times10^{-3}$&unstable \\
  $10^8$ & 1000 & $4\times10^4$ & 0.0625  & 0.025 & 25 &1.08 & $9.09\times10^{-3}$ &1.5 & $1.44\times10^{-5}$&stable \\
  $10^8$ & 1000 & $5\times10^4$ & 0.04  & 0.02 & 20 & 1.09 & $7.61\times10^{-3}$ & 1.5 & $1.10\times10^{-5}$&stable \\
  \hline
  $10^8$ & 5000 & 5000 & 4.00  & 1.00 & 5000 & 0.89 & $1.79\times10^{-1}$ & 3.67 & $1.76\times10^{-2}$&unstable \\
  $10^8$ & 5000 & $10^4$ & 1.00  & 0.50 & 2500 & 0.86 & $3.19\times10^{-2}$ & 1.5 & $3.38\times10^{-4}$&stable \\
  $10^8$ & 5000 & $2\times10^4$ & 0.25  & 0.25 & 1250 & 0.85 & $1.21\times10^{-2}$ & 1.5 & $1.63\times10^{-4}$&stable \\
  \hline
  $10^8$ & $10^4$ & $10^4$ & 1.00  & 1.00 & $10^4$ & 0.22 & $1.11\times10^{-1}$ & 1.5 & $5.10\times10^{-4}$&stable \\
  $10^8$ & $10^4$ & $2\times10^4$ & 0.25  & 0.50 &  5000 & 0.74 & $1.25\times10^{-2}$ & 1.5 & $2.43\times10^{-4}$&stable \\
  \hline 
  $10^9$ & 1000 & $10^4$ & 10.00  & 0.10 & 100 & 0.85 & $3.63\times10^{-2}$ & 5.64 & $1.91\times10^{-1}$&unstable \\
  $10^9$ & 1000 & $4\times10^4$ & 0.625  & 0.025 & 25 & 1.14 & $8.87\times10^{-3}$ & 2.69 & $9.82\times10^{-3}$&unstable \\
  $10^9$ & 1000 & $5\times10^4$ & 0.40  & 0.02 & 20 & 1.14 & $7.87\times10^{-3}$ & 1.92 & $5.16\times10^{-3}$&unstable \\
  $10^9$ & 1000 & $10^5$ & 0.10  & 0.01 & 10 & 1.15 & $5.98\times10^{-3}$ & 1.91 & $1.14\times10^{-3}$&unstable \\
  \hline
  $10^9$ & 5000 & $2\times10^4$ & 2.50  & 0.25 & 1250 & 0.96 & $2.44\times10^{-2}$ & 3.20 & $1.00\times10^{-2}$&unstable \\
  $10^9$ & 5000 & $3\times10^4$ & 1.11  & 0.17 & 833 & 0.39 & $2.43\times10^{-2}$ & 2.56 & $4.37\times10^{-3}$&unstable \\
  $10^9$ & 5000 & $4\times10^4$ & 0.625  & 0.125 & 625 & 0.97 & $1.19\times10^{-2}$ & 1.95 & $6.91\times10^{-4}$&unstable \\
  $10^9$ & 5000 & $5\times10^4$ & 0.40  & 0.10 & 500 & 0.96 & $5.98\times10^{-3}$ & 1.50 & $7.08\times10^{-5}$&stable \\
  \hline
  $10^9$ & $10^4$ & $1.5\times10^4$ & 4.44  & 0.67 & 6670 & 0.80 & $4.57\times10^{-2}$ & 3.20 & $1.27\times10^{-2}$&unstable \\
  $10^9$ & $10^4$ & $2\times10^4$ & 2.50  & 0.50 & 5000 & 0.82 & $3.54\times10^{-2}$ & 2.56 & $5.74\times10^{-3}$&unstable \\
  $10^9$ & $10^4$ & $3\times10^4$ & 1.11  & 0.33 & 3333 & 0.83 & $1.59\times10^{-2}$ & 1.50 & $9.81\times10^{-4}$&stable \\
  $10^9$ & $10^4$ & $5\times10^4$ & 0.40  & 0.20 & 2000 & 0.84 & $9.82\times10^{-3}$ & 1.50 & $1.96\times10^{-4}$&stable \\
  \hline
  $10^{10}$ & $10^4$ & $2\times10^4$ & 25.00  & 0.50 & 5000 & 1.07 & $1.44\times10^{-2}$ & 7.94 & $6.28\times10^{-2}$&unstable \\
  $10^{10}$ & $10^4$ & $5\times10^4$ & 4.00  & 0.20 & 2000 & 0.99 & $5.95\times10^{-3}$ & 2.95 & $6.08\times10^{-3}$&unstable \\
  $10^{10}$ & $10^4$ & $10^5$ & 1.00  & 0.10 & 1000 & 0.42 & $5.05\times10^{-3}$ & 2.71 & $1.14\times10^{-3}$&unstable \\
  \hline
  $10^{11}$ & $10^4$ & $10^5$ & 10.00  & 0.10 & 1000 & 1.12 & $2.68\times10^{-3}$ & 5.42 & $1.51\times10^{-2}$&unstable \\
  $10^{11}$ & $10^4$ & $10^6$ & 0.10  & 0.01 & 100 & 0.47 & $7.00\times10^{-4}$ & 5.40 & $1.87\times10^{-2}$&unstable \\
  \end{tabular}}
  \end{center}
  \caption{Summary of the simulations performed at $L_z=30$. The meaning of the non-dimensional groups and the definitions of the flow characteristics are in the text. The integral characteristics $E$, $\overline{T}$, and $E_y$ are computed for fully-developed flows using volume- and, in the case of unstable flows, time-averaging. The maximum difference in vertical velocity ${\Delta u_z}^{max}$ in unstable flows is computed at the moment of the strongest jet growth just before the breakdown.}
  \label{tab2}
\end{table*}

One can hypothesize that the stability of the flow is primarily determined by the two parameter groups representing the strengths of the buoyancy and magnetic damping forces in relation to the inertial force (see (\ref{me})): $\Gras/\Rey^2$ and $\Ha/\Rey$. The summary of all the computed flow regimes shown in Fig.~\ref{fig4} (see also table \ref{tab2}) confirms this hypothesis. In fact, the data indicate that the stability can be determined by the ratio between the two groups
\begin{equation}
\label{pipi}
\Pi\equiv \frac{\Gras}{\Ha\Rey}
\end{equation}
with the stability threshold at about $\Pi_{crit}=4$.

The flow patterns are simple and do not differ much from each other in the stable regime. On the contrary, in the unstable regime, the patterns of the developed flows are complex and diverse. The properties of such flows, cannot be reliably identified by $\Pi$ or even by the two groups $\Gras/\Rey^2$ and $\Ha/\Rey$. All three independent parameters $\Gras$, $\Rey$, and $\Ha$ are significant here. This is clearly demonstrated in Fig.~\ref{fig5}, showing three unstable flows with strongly different sets of $\Gras$, $\Rey$, and $\Ha$, but the same combinations  $\Gras/\Rey^2=4$ and $\Ha/\Rey=0.2$. 

The flow structures found for the unstable regime can be roughly classified into three types.
One type has already been illustrated in Figs.~\ref{fig2}d-e and \ref{fig3}. The secondary rolls developing in the result of the instability are weak.
The integrity of the jets is maintained. 
As illustrated in Fig.~\ref{fig2}f, the instability evolves in the shear layer between the upward and downward jets and does not destroy the jet pattern. In particular, the thin upward jet near the heated wall exists throughout the breakdown events. The integral properties are nearly constant with small amplitude of fluctuations.

Another type of the unstable flow structure is illustrated in Fig.~\ref{fig6}. The flow has much stronger rolls resulting in strong mixing, so the jets are destroyed from time to time. The rolls penetrate far from the heated wall and may even reach the opposite wall creating hot spots near it. The main time period and amplitude of the oscillation of the temperature signal are much larger than in the unstable flows of the first type.
The typical flow evolution is illustrated in Fig.~\ref{fig7}. It can be viewed as a cycle consisting of the periods of growth of high-energy jets interrupted by large-scale instability events leading to development of strong mixing zones and the jets' complete breakdowns. The mixing zones are transported downwards by the mean flow. Locally, the evolution manifests itself as the quasi-periodic evolution of temperature shown in Fig.~\ref{fig6}e. The periods of growth and sharp drops of temperature correspond, respectively, to the periods of jet growth and the moments when the mixing zones pass the selected location.

Flows at high $\Gras$ are likely to take yet another form illustrated in Fig.~\ref{fig8}. Instead of one long upward jet as in the previous two forms, there are several shorter thin jets along the heated wall. 
In other respects, the flow reminds the flow of the second type. In particular, it demonstrates the quasi-periodic evolution consisting of growth and complete breakdown of upward jets. The smaller axial wavelength means that the typical period of local temperature fluctuations is smaller than in the second-type flows (compare the signals in Fig.~\ref{fig8}e and Fig.~\ref{fig6}e). The non-dimensional amplitude of the temperature fluctuations is also smaller, but still significant.

We have already mentioned that the flow patterns are not solely determined by the two groups $\Gras/\Rey^2$ and $\Ha/\Rey$, but depend on all three parameters $\Gras$, $\Rey$, and $\Ha$. We now analyze the effects of individual parameters on the flow.

The effect of the Reynolds number is a product of three competing mechanisms. 
Firstly, the value of $\Rey$ affects the strength of the buoyancy force through $\Gras/\Rey^2$. Decrease in $\Rey$ means stronger buoyancy effect and stronger instability. At the same time, the value of $\Rey$ affects the strength of the magnetic damping through $\Ha/\Rey$. Decrease in $\Rey$ implies stronger damping effect  and, thus,  may lead to lower flow energy and possible stabilization. Finally, decrease in $\Rey$ also means stronger viscous dissipation and heat conduction, i.e. tendency to decrease of kinetic energy and stabilization. Considering the large values of $\Rey$ and $\Gras$ explored in our study, one can assume that the influence of viscosity and conductivity  and of their variation is weak in comparison to the other two effects mentioned above.  

The data in table \ref{tab2} clearly show that one tendency, namely the effect of $\Rey$ on the buoyancy force, dominates the other two. We see in the table that increase of $\Rey$ at constant $\Gras$ and $\Ha$ invariably leads to reduction of $\Delta u_z^{max}$ (weaker upward jet) and $E_y$ (lower amplitude of rolls developing in the result of the instability) and, finally, stabilization of the flow. As an illustration of the effect, Fig.~\ref{fig9} shows the unstable flows at $\Gras=10^8$, $\Ha=1000$ and three values of $\Rey$: 5000, $2\times 10^4$ and $4\times 10^4$. The weakening of the jets ( $\Delta u_z^{max}$ changes from 5.04 to 2.22) and the change of the instability type to a less intensive one is clearly visible in the first two plots. The third plot shows a stable flow.

The effect of Re can be given a simple physical explanation based on the net energy balance in the flow. The unstable stratification is a result of the balance between the heat flux at the wall and the convective heat transfer in the downward direction by the mean axial velocity. Integration of the steady-state heat equation reveals the textbook fact that the energy balance in this case results in the mean temperature growing linearly downstream with the slope $(\Rey\Pran)^{-1}$. Increase of $\Rey$ reduces the stratification, thus reducing the buoyancy force and the strength of the jets, which in turn suppresses the instability.

The effect of larger $\Ha$ is that of stronger magnetic damping. This implies reduction of the flow's kinetic energy and suppression of the instability. This effect is illustrated in Fig.~\ref{fig10}. Flows at $\Gras=10^9$, $\Rey=2\times10^4$, and $\Ha=5000$ and $\Ha=10^4$ are compared. The flows are both unstable, but they have different evolutions. It takes longer time for the flow at higher $\Ha$ to reach the fully developed state.
In the fully developed state, increase of $\Ha$ results in weaker fluctuations, lower average kinetic energy and higher mean temperature.

The effect of $\Gras$ is associated with the buoyancy effect and straightforward. Increase in $\Gras$ means stronger buoyancy force, which leads to stronger growth of the upward jet. Stronger instability is expected when the two-jet structure breaks down. 
This effect is illustrated in Fig.~\ref{fig11}.
Figs.~\ref{fig11}a, b show the two flows at $\Ha=10^4$, $\Rey=2\times10^4$, and $\Gras=10^8$ and $10^9$. We see the destabilization of the flow at higher $\Gras$. Figs.~\ref{fig11}c, d illustrate the effect of $\Gras$ on unstable flows on the example of the two cases: $\Ha=10^4$, $\Rey=5\times10^4$, and $\Gras=10^{10}$ and $10^{11}$. We see that $\Gras$ has significant influence in the flow structure. The flow at $\Gras=10^{11}$ has stronger instability in the form of  convection structures with shorter typical axial wavelength. 

The results discussed so far are for the non-dimensional duct length $L_z=30$. It would be interesting to know whether the duct length significantly affects the flow behavior. 
The results of our tests are illustrated in Fig.~\ref{fig12} by the flows at $\Gras=10^9$, $\Ha=5000$ and $\Rey=2\times10^4$, computed in the ducts with lengths $L_z=30$ and $L_z=60$. 
Both the flows are unstable, but increasing domain length changes the structure of the flow.
Most noticeably, the typical length, to which the upward jet grows before experiencing a breakdown increases to $\sim 20$ in the longer domain. This is comparable to the length of the heated part of the duct in the shorter domain. So the domain length and the presence of the buffer layers limit the development of the flow structures in the shorter domain simulations. While making this conclusion, we note that the constraint is, to a degree, analogous to the real geometry constraints in the fusion reactor blankets. In the current designs, the typical poloidal ducts have width $10$ to $20\,cm$ and length $1$ to $2 \,m$ \cite{Abdou:2015}.

Finally, we consider implications of our results for the fusion reactor blankets, where the main interest is the transport and mixing properties of the flow and the possibility of potentially destructive features of the temperature field: large-amplitude fluctuations, strong spatial gradients, and distinct hot and cold spots. Firstly, we note that in practically the entire range of the parameters corresponding to the typical conditions of a blanket ($10^{9}\le\Gras\le 10^{12}$, $\Ha=10^4$, $10^4\le\Rey\le10^5$), the values of $\Pi$ (see \eqref{pipi}) are larger or much larger than the stability threshold $\Pi_{cr}\approx 4$. The flows are expected to be unstable. All three types of the unstable flow behavior can be observed depending on the specific duct's parameters. Our results imply that there will be no hot and cold spots in the straight portion of the duct (although such spots may develop in a real duct near the inlet and exit manifolds). The degree of mixing within the flow is expected to vary between moderate (for the flows of the first type) to very strong (for the flows of the third type). This conclusion may seem trivial, but it is in a strong contrast with practically zero mixing in a laminar flow that would be found in the duct at such high $\Ha$ if the thermal convection were not taken into account (as discussed, e.g. in \cite{Zikanov:2014}, turbulence does not exist in an isothermal flow at $\Ha/\Rey > 0.005$).

Most importantly, the thermal convection in the vertical ducts with downward flow would practically inevitably lead to large-amplitude fluctuations of temperature. As an example, we consider the flow at $\Gras=10^{10}$, $\Ha=10^4$, $\Rey=5\times10^4$. The signal of temperature at a point  close to the heated wall and the  spectral decomposition of the signal are shown in Fig.~\ref{fig13}. The dominating frequency $\sim0.08$ (see Fig.~\ref{fig13}b) corresponds to the typical period of oscillations $\sim12.5$ which can be observed in Fig.~\ref{fig13}a. The typical amplitude of the temperature fluctuations is considerable. If we convert the solution into dimensional units using the duct half-width $d=10\,cm$, mean velocity $10\,cm/s$, and the physical properties of PbLi at $570\,K$, we find the fluctuation amplitude of $37\,K$. In a narrower duct with $d=5\,cm$, the similar estimate is $296\,K$. Even higher amplitudes are predicted by our simulations at higher $\Gras$. The two main conclusions from the estimates are that the fluctuations present a potentially very serious problem for the blanket design and that the problem may need to be revisited in the framework of the non-Boussinesq and perhaps even multiphase model.

\section{Conclusions}
\label{conclusions6}

The study of thermal convection in a downward flow in a vertical duct with one wall heated and a strong transverse magnetic field reveals that there are two states of the flow: stable and unstable. The stability threshold can be identified  by the parameter $\Gras/(\Ha\Rey)$ with the threshold value around $4$. In the stable flows, the velocity profiles become asymmetric due to the action of the buoyancy force. The unstable flows all develop thin reverse (upward) jet near the heated wall. The shear layer between the jet and the downward flow is unstable to two-dimensional rolls of Kelvin-Helmholtz type.
The parametric analysis has shown that the unstable flows have different structures depending on the values of $\Gras$, $\Ha$ and $\Rey$. Three distinctive types of the structure have been identified. In many cases the instability leads to quasi-periodic breakdown of the jets, which in turn results in high-amplitude low-frequency oscillations of temperature.

Qualitatively similar oscillations are observed in the experiments \cite{Kirillov:2016}. The experimental flow cannot be reproduced in the framework of our model because at the parameters of the experiment ($\Gras\approx 10^8$, $\Rey\approx 10^4$, $\Ha\le 800$) the conditions of the two-dimensional approximation \eqref{asl} are not satisfied to a sufficient degree of certainty. Future three-dimensional simulations of the entire experimental test section similar to those performed in \cite{Zikanov:2016} are warranted.

\begin{acknowledgments}
The authors are thankful to Dmitry Krasnov for continuing help with code development and to Yaroslav Listratov and Ivan Belyaev for stimulative discussions in the course of the work.
 Financial support was provided by the US NSF (Grant CBET 1435269).
 \end{acknowledgments}

\bibliographystyle{plain} 


\end{document}